\newif\iftwocol

\iftwocol
    \documentclass[10pt]{iopart}
\else
    \documentclass[12pt]{iopart}
\fi

\usepackage{iopams}
\usepackage{graphicx}

\expandafter\let\csname equation*\endcsname\relax
\expandafter\let\csname endequation*\endcsname\relax

\usepackage{amsmath}
\usepackage{amssymb}

\begin{document}

\title[GPR Resolution Limits]{Quantifying Resolution Limits in Pedestal Profile Measurements with Gaussian Process Regression}

\author{Norman M. Cao\(^{1}\), David R. Hatch\(^{1,5}\), Craig Michoski\(^{2,5}\), Todd A. Oliver\(^{2}\), David Eldon\(^{3}\), Andrew Oakleigh Nelson\(^{4}\) and Matthew Waller\(^{5}\)}

\address{\(^{1}\)Institute for Fusion Studies, The University of Texas at Austin, Austin, TX 78712, United States of America}
\address{\(^{2}\)Oden Institute for Computational Engineering and Sciences, The University of Texas at Austin, Austin, TX 78712, United States of America}
\address{\(^{3}\)General Atomics, San Diego, CA 92121, United States of America}
\address{\(^{4}\)Columbia University, New York, NY 10027, United States of America}
\address{\(^{5}\)Sophelio LLC, Austin, TX 78759, United States of America}
\ead{norman.cao@austin.utexas.edu}
\vspace{10pt}
\begin{indented}
\item[]\today
\end{indented}

\begin{abstract}
Edge transport barriers (ETBs) in magnetically confined fusion plasmas, commonly known as pedestals, play a crucial role in achieving high confinement plasmas.
However, their defining characteristic, a steep rise in plasma pressure over short length scales, makes them challenging to diagnose experimentally.
In this work, we use Gaussian Process Regression (GPR) to develop first-principles metrics for quantifying the spatiotemporal resolution limits of inferring differentiable profiles of temperature, pressure, or other quantities from experimental measurements.
Although we focus on pedestals, the methods are fully general and can be applied to any setting involving the inference of profiles from discrete measurements.
First, we establish a correspondence between GPR and low-pass filtering, giving an explicit expression for the effective `cutoff frequency' associated with smoothing incurred by GPR.
Second, we introduce a novel information-theoretic metric, \(N_{eff}\), which measures the effective number of data points contributing to the inferred value of a profile or its derivative.
These metrics enable a quantitative assessment of the trade-off between `over-fitting' and `over-regularization', providing both practitioners and consumers of GPR with a systematic way to evaluate the credibility of inferred profiles with respect to their degree of regularization.
We apply these tools to develop practical advice for using GPR in both time-independent and time-dependent settings, and demonstrate their usage on inferring pedestal profiles using measurements from the DIII-D tokamak.
\end{abstract}

%
%
%
%

\iftwocol
    \ioptwocol
    \newcommand{\halfcolwidth}{\linewidth}
\else
    \newcommand{\halfcolwidth}{0.5\linewidth}
\fi
%

\newcommand{\norm}[1]{\left\lVert#1\right\rVert}
\newcommand{\ev}[1]{\left\langle#1\right\rangle}
\newcommand{\dd}{\mathrm{d}}

\section{Introduction}

A major goal in magnetic confinement fusion energy is the development of predictive models of transport for energy, particles, and momentum across transport barriers.
Edge transport barriers (ETBs), generally known as pedestals, are of particular interest due to their critical role in the performance of H-mode and other enhanced confinement regimes \cite{wagner_quarter-century_2007,groebner_elements_2023}.
Transport across pedestals must balance good confinement in the plasma core with a multitude of competing requirements such as heat flux management to the divertor and impurity accumulation, constituting the so-called ``core-edge integration'' challenge \cite{dux_accumulation_2003,pitts_physics_2019,viezzer_prospects_2023}.
Although ETBs are the primary focus of this work, internal transport barriers (ITBs) have also been explored as ways to improve the performance of fusion devices and meet these challenges \cite{ida_internal_2018}, and the tools developed here are applicable to them as well.

Achieving reliable control and quantification of trade-offs of transport in pedestals and other transport barriers requires experimentally validated models of physical processes in these regions \cite{terry_validation_2008,greenwald_verification_2010,fischer_estimation_2020}.
However, the very property which makes transport barriers attractive for confinement, their steep rise in plasma pressure over a short length scale, makes them challenging to diagnose experimentally.
The typical radial width of a pedestal \(\ell\) is only a small fraction of the device size \(L\), with \(\ell/L \sim 2\text{--}10\%\).
For diagnostics such as Thomson scattering, only a few channels might cover the width of the pedestal, limiting the accuracy to which profile gradients can be inferred from experimental measurements.
Meanwhile, transport mechanisms frequently exhibit critical gradient phenomena where turbulent fluxes become extremely sensitive to gradients of density, temperature, or pressure above certain instability thresholds.
Transport mechanisms are also sensitive to velocity shear, which can act to stabilize or entirely quench transport.
In cases where the radial electric field is dominated by diamagnetic flow \(E_r \sim \nabla p\), computing \(E \times B\) shear requires accurately inferring the second derivative of pressure profiles from measurements.

Further complicating matters in pedestal modeling is the interest in understanding temporally-dependent phenomena such as the edge-localized mode (ELM) cycle \cite{laggner_inter-elm_2019,diallo_review_2020} or the L-H transition \cite{schmitz_role_2012}, which can involve fast temporal evolution of pedestal profiles.
These time-dependent phenomena relate to problems such as the maximum pedestal pressure which can be achieved after an ELM, as well the physics of the minimum power threshold necessary for achieving H-mode.
Beyond pedestals, there is growing interest in direct profile measurements of mesoscale phenomena, such as `staircase' profiles in L-mode and H-mode \cite{dif-pradalier_finding_2015,ashourvan_formation_2019,liu_evidence_2021,choi_mesoscopic_2024}, as well as profile variation due to magnetic islands \cite{waelbroeck_theory_2009,choi_interaction_2021}.

In this work we address the problem of uncertainty quantification in inferences of differentiable profiles of plasma properties from spatially localized measurements.
The key question we address is: what is the maximum spatial and temporal scale of features which can be reliably inferred from a given set of profile measurements?
To make this problem quantitative, we adopt the popular approach of Gaussian Process Regression (GPR) \cite{rasmussen_gaussian_2008,chilenski_improved_2015,ho_application_2019,pavone_machine_2023,michoski_gaussian_2024}.
GPR is enormously flexible, being used to perform principled inference and uncertainty quantification from complex experimental setups \cite{chilenski_experimentally_2017,cao_hysteresis_2019,nishizawa_plasma_2021,kwak_bayesian_2022}, produce tomographic inversions \cite{li_bayesian_2013,langenberg_forward_2016,wang_bayesian_2019,chao_gaussian_2020,matos_deep_2020}, and provide surrogate models of complex phenomena \cite{preuss_gaussian_2016,rodriguez-fernandez_nonlinear_2022}.

GPR has two important advantages: it does not impose a fixed functional form for the prior distribution of functions which can be inferred, and it can provide statistically principled error quantification for both the inferred profiles and their derivatives.
This can be contrasted with `parametric' regression techniques, such as spline fits or the modified hyperbolic tangent functions commonly used in pedestal modeling \cite{groebner_scaling_1998,groebner_progress_2001}, which constrain the inferred profiles to lie within a predetermined family of functions characterized by a finite number of parameters.
While GPR is often called `non-parameteric', we discuss how one of the main challenges of interpreting results from GPR is the problem of kernel and kernel hyperparmeter selection, as these choices often make significant impacts on the resulting inferred profiles.
In particular, we discuss how pushing the resolution limits using GPR necessarily involves a trade-off between `over-regularization' and `over-fitting'.

To tackle these issues, this paper presents two key results.
The first is the presentation of a mathematical correspondence between GPR and low-pass filtering using an `equivalent kernel' analysis, which provides an unambiguous interpretation for the hyperparameters used in GPR.
We give an explicit relationship between the length scale parameter of the Mat\'{e}rn family of kernels, which includes the popular squared exponential kernel as a special case, and the cutoff frequency associated with this low-pass filter.
In particular, we show that the relationship between the length scale parameter and (inverse) cutoff frequency also depends on the signal-to-noise rate, implying that the intuitive choice of taking the length scale parameter equal to some characteristic physical length scale does not necessarily apply the right degree of regularization (i.e. smoothing) to the profiles.

The second key result is the definition of a conveniently computed metric called \(N_{eff}\) which quantifies the effective number of measurements which contribute to the inferred value of the fitted profile or its derivative at a given location.
Roughly speaking, this quantity is equal to
\begin{equation*}
	N_{eff} = \frac{\text{Total information from measurements}}{\text{Weighted average information per measurement}}
\end{equation*}
Rather than identify a single best choice of kernel, \(N_{eff}\) provides an interpretable information-theoretic metric useful for detecting when a given kernel is vulnerable to over-fitting.
Together, these two results provide tools to replace trial-and-error during the fitting process or hyperprior development process with quantitative metrics that can be used to engineer the tradeoff between over-fitting and over-regularization -- something that simpler parametric fits cannot provide.

This paper is organized as follows:
We begin in section \ref{sec:setup} by introducing notation and conventions for Gaussian Process Regression adopted in this work, and discuss trade-offs in GPR methodology.
Afterwards, readers who are interested in applications of the results can skip ahead to section \ref{sec:d3d}, where we demonstrate their usage in a practical situation involving pedestal profiles on DIII-D.
Otherwise, we follow in section \ref{sec:lowpass} with a discussion of how GPR can be viewed as a form of low-pass filtering and demonstrate the insufficiency of the kernel length scale in determining the length scale associated with this filtering.
In section \ref{sec:neff} we derive an expression for \(N_{eff}\) and provide several main properties it satisfies.
In section \ref{sec:d3d} we demonstrate the usage of these tools in practical situations demonstrating how they affect profile fits.
Finally, in section \ref{sec:summary} we summarize the work and discuss further potential applications enabled by the results.

\section{Profile Fitting with GPR} \label{sec:setup}

In this section, we introduce some background on GPR and discuss how the choice of GPR methodology can influence the resulting profiles inferred from the measurements.
In section \ref{subsec:setup_overview} we fix notation and give a conceptual overview of Gaussian Processes and the relation to GPR.
Then in section \ref{subsec:setup_hyperparameters} we introduce the problem of hyperparameter selection in GPR, in particular focusing on the trade-off between over-regularization and over-fitting inherent to all profile fitting techniques.

\subsection{Conceptual Overview of GPR} \label{subsec:setup_overview}

We are interested in studying the following problem:
Suppose \(f(x)\) is an unknown function on a domain \(D\) and we have \(N\) noisy measurements \(\mathbf{Y} = [Y_1, ..., Y_N]^T\) of this profile at points \(\mathbf{x} = [x_1, ..., x_N]^T\) given by
\begin{equation}
	Y_i = f(x_i) + \varepsilon_i
\end{equation}
where \(\boldsymbol{\varepsilon} = [\varepsilon_1, ..., \varepsilon_N]^T\) is a vector of random variables which capture the deviation of the measurements from \(f(x)\).
Typically \(f\) models some `mean' or `ensemble average' profile and \(\varepsilon_i\) models random measurement noise, but the latter can also capture deviations of the measurements from \(f(x)\) due to time-dependent fluctuations and systematic uncertainties.
In general, \(x\) can be either a scalar quantity for 1D profiles, or can be a vector quantity for multivariate profiles.

In general, functions have an infinite number of degrees of freedom, whereas the number of measurements we have is decidedly finite.
Hence, the problem of inferring \(f(x)\) from measurements is in some sense always severely underdetermined, so we must always make some choice about how to `fill in the gaps' between measurements.
Gaussian Process Regression (GPR), also known as kriging in the literature, is a particular choice of how to resolve this severe underdetermination which has the nice properties of (1) not imposing a fixed functional form for the prior distribution of functions which can be inferred, and (2) having a principled way to compute statistical uncertainties for the resulting function \(f(x)\) and any of its derivatives \cite{rasmussen_gaussian_2008}.

Several works give extensive discussions of GPR and how to apply it in the context of profile fitting \cite{chilenski_improved_2015,michoski_gaussian_2024}, so here we will give a short overview of GPR and the key results we will use.
Conceptually, a \textit{Gaussian process} (GP) is a way of assigning a probability measure to some ensemble (i.e. set) of functions \(H\), with each realization \(f \in H\) being a different possible function \(f(x)\).
Similar to how the distribution of a Gaussian random vector is entirely specified by the mean and covariance matrix of its components, a GP is entirely specified by a mean \(\mu(x)\) and covariance kernel \(k(x,x')\), which satisfy
\begin{gather*}
    \mathbb{E}[f(x)] = \mu(x) \\
    \operatorname{Cov}[f(x),f(x')] = k(x,x')
\end{gather*}
Typically the ensemble \(H\) can be extended to include all functions with square-integrable derivatives up to some given order, i.e. a Sobolev space.

In the following, we denote \(k(\mathbf{x},x)\) to be the column vector with components \(k(x_i,x)\), \(k(x,\mathbf{x})\) to be the row vector with components \(k(x,x_j)\), and \(k(\mathbf{x},\mathbf{x})\) to be the \(N \times N\) symmetric matrix with components \(k(x_i,x_j)\).
Notably for any finite collection of points \(\mathbf{x}\), the vector \(f(\mathbf{x}) := [f(x_1),...,f(x_N)]^T\) will be distributed as a multivariate Gaussian random variable with mean \(\mu(\mathbf{x}) := [\mu(x_1),...,\mu(x_N)]^T\) and covariance \(k(\mathbf{x},\mathbf{x})\).

GPR can be thought of as a way to update the mean and covariance of a GP to match a set of noisy measurements \(\mathbf{Y}\).
In practice, GPR begins by specifying a prior mean \(\mu(x;\boldsymbol{\theta})\) and covariance kernel \(k(x,x';\boldsymbol{\theta})\), usually depending on some set of hyperparameters \(\boldsymbol{\theta}\).
In a Bayesian view, this prior GP encodes our assumptions about how `likely' it is to observe a given realization \(f \in H\) of our ensemble.
For brevity we will typically suppress the dependence on the hyperparameters \(\boldsymbol{\theta}\).

GPR is based on the fact that if \(\boldsymbol{\varepsilon}\) is modeled as a multivariate Gaussian with mean zero and covariance matrix \(\boldsymbol{\Sigma}_{\varepsilon}\), then the posterior distribution of \(f\) conditioned on knowledge of \(\mathbf{Y}\) will also be a GP.
The posterior mean, defined by the conditional expectation \(\hat{f}(x) := \mathbb{E}[f(x) | \mathbf{Y}]\), has an explicit expression
\begin{equation} \label{eq:fhat}
\begin{gathered}
    \hat{f}(x) = \mu(x) + \boldsymbol{\beta}^T(x) (\mathbf{Y} - \mu(\mathbf{x})) \\
    \boldsymbol{\beta}(x) := \left[k(\mathbf{x}, \mathbf{x}) + \boldsymbol{\Sigma}_{\varepsilon}\right]^{-1} k(\mathbf{x},x)
\end{gathered}
\end{equation}
where \(\boldsymbol{\beta}(x) = [\beta_1(x), ..., \beta_N(x)]^T\) is the vector of so-called `regression coefficients'.
Observe that the number of regression coefficients grows as the number of data points grows, which allows the inferred profile to become more flexible as more data is added.
The posterior covariance, defined as the conditional covariance \(\hat{k}(x,x'):= \operatorname{Cov}[f(x),f(x')|\mathbf{Y}]\), also has an explicit expression
\begin{equation*}
    \hat{k}(x,x') = k(x,x') - k(x,\mathbf{x}) \left[k(\mathbf{x},\mathbf{x}) + \boldsymbol{\Sigma}_{\varepsilon}\right]^{-1} k(\mathbf{x},x')
\end{equation*}

In practice, the `fit' of the data \(\mathbf{Y}\) by GPR is usually taken to be the posterior mean \(\hat{f}(x)\), with the posterior covariance \(\hat{k}(x,x')\) determining the error bars.
We will refer to \(\hat{f}(x)\) as the \textit{inferred} profile.
In a frequentist view, the meaning of this inferred profile is that if we were to sample \(f \in H\) based on our prior GP, then weight them by the probability that they are consistent with the measurements \(\mathbf{Y}\), then the weighted mean and covariance of these samples would be given by \(\hat{f}\) and \(\hat{k}\) respectively.

The derivative of the inferred profile can be explicitly computed as
\begin{equation} \label{eq:fhat_deriv}
	\hat{f}'(x) = \mu'(x) + \boldsymbol{\beta}'^T(x) (\mathbf{Y} - \mu(\mathbf{x}))
\end{equation}
which satisfies \(\hat{f}'(x) = \mathbb{E}[f'(x) | \mathbf{Y}]\), assuming the kernel \(k\) is sufficiently differentiable.
There is also an explicit formula for the covariance of the derivatives \(f'\).

For concreteness, for the prior we will focus on the family of stationary Mat\'{e}rn kernels \cite{rasmussen_gaussian_2008}.
In 1D, these kernels are given by \(k(x,x') = \kappa(x-x')\) with
\begin{equation} \label{eq:matern}
    \kappa(x) := \begin{cases}
        \sigma^2 \frac{2^{1-\nu}}{\Gamma(\nu)}\left(\sqrt{2 \nu} \frac{|x|}{\ell}\right)^{\nu} K_\nu\left(\sqrt{2\nu} \frac{|x|}{\rho}\right) & 0 < \nu < \infty \\
        \sigma^2 \exp\left(-\frac{x^2}{2\ell^2}\right) & \nu = \infty
    \end{cases}
\end{equation}
where \(\Gamma\) is the gamma function, and \(K_\nu\) is the modified Bessel function of the second kind.
This family is parameterized by three hyperparmeters: \(0 < \nu \le \infty\) is the smoothness parameter, \(\sigma^2 > 0\) is the variance parameter, and \(\ell > 0\) is the length scale parameter.
For \(\nu = \infty\) it is equivalent to the squared exponential kernel, which is commonly used in profile fitting.

\subsection{Hyperparameter Selection} \label{subsec:setup_hyperparameters}

In this section, we outline how hyperparameters affect the behavior of GPR inferences and discuss potential issues that can arise if they are not properly chosen.
We begin by providing some intuition for the Mat\'{e}rn kernel hyperparameters.
Then, we summarize several common methods for choosing or marginalizing over hyperparmeters, and discuss differences in the uncertainty estimates and interpretations of inferences from these methods.
Finally, we discuss the need for quantitative metrics to assess whether a chosen set of hyperparameters lead to over-regularization or over-fitting, giving criteria to determine if an inference represents a credible interpretation of the data.
In this work we remain agnostic to how the hyperparameters are chosen or marginalized over, as the methods developed in this work focus on how to interpret any set of hyperparameters from any method.
For more detailed prescriptions for hyperparameter inference, readers are directed to the references \cite{chilenski_improved_2015,michoski_gaussian_2024}.

Recall that the kernel \(k\) determines the prior distribution of functions \(f\) over which the GPR inference is carried out.
Thus, different choices of hyperparameter encode different assumptions about what the `typical' \(f\) looks like.
The easiest hyperparameter to understand is the variance parameter \(\sigma^2\), which determines the typical amplitude of \(f(x)\) via the relation
\begin{equation*}
    \mathbb{E}[(f(x) - \mu(x))^2] = \sigma^2
\end{equation*}
It is also clear from this that \(\sigma^2\) has the same dimensions as \(f^2\).
Next, the smoothness hyperparameter \(\nu\), which is dimensionless, controls the differentiability of the realizations \(f \in H\).
In general \(f\) will be \(\lceil \nu \rceil - 1\) times differentiable in the mean-square sense \cite{rasmussen_gaussian_2008}.
Finally, the length scale parameter \(\ell\) has the same dimensions as the measurement variable \(x\).

\begin{figure}
    \centering
    \includegraphics[width=\halfcolwidth]{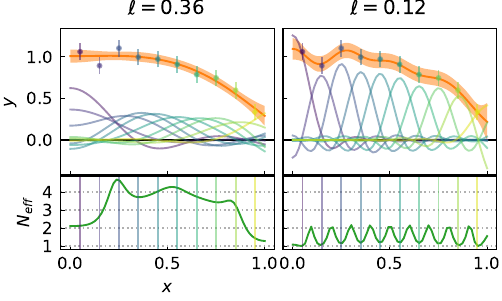}
    \caption{(Top) Illustration of how the length scale \(\ell\) parameter affects profiles (orange curve with \(\pm34\%\) confidence intervals) inferred using GPR. The decomposition of this profile into scaled regression coefficient functions \(Y_i \beta_i(x)\) (smooth curves, various colors) associated with the data points \(Y_i\) (scattered points, various colors) is also shown. (Bottom) Plot of the \(N_{eff}(x)\) metric defined in section \ref{sec:neff}. The colored vertical lines show the \(x\) location of the data points. \(N_{eff} \approx 1\) indicates only one measurement is contributing to the inference of the profile at a given point, indicating overfitting in the \(\ell=0.12\) case.}
    \label{fig:gpr_basis}
\end{figure}

Generally speaking, for fixed values of \(\sigma^2\) and \(\nu\), as the length scale \(\ell\) is decreased, the inferred profile picks up on smaller-scale fluctuations present in the data \(\mathbf{Y}\).
An example of this is shown in figure \ref{fig:gpr_basis}, which illustrates how two example inferred profiles using the same data are decomposed into regression coefficients \(\beta_i(x)\) for different values of \(\ell\).
Notice that for the case with larger \(\ell\), the functions \(\beta_i(x)\) are broader, which smooth out fluctuations in the data \(\mathbf{Y}\).
Meanwhile, for the case with smaller \(\ell\), the \(\beta_i(x)\) are narrower, which lead to larger fluctuations in the inferred profile.
As we will see later in section \ref{sec:lowpass}, this difference in behavior explicitly reflects the bias that the assumptions on \(f\) introduce into the GPR estimate \(\hat{f}\).

Given the influence of the hyperparameters on the inferred profile in GPR, care must be taken in how to specify the hyperparameters.
The error bars shown in figure~\ref{fig:gpr_basis} are rigorous and exact within the assumptions of the uncertainty model.
However, any calculation is only as good as its model!
A fully statistically principled approach requires a more careful examination of the assumptions underlying the inference process.
Here, we first outline the approaches by which hyperparameters are usually determined, then discuss necessary conditions for the physical credibility of the inferences produced by these approaches.

Broadly speaking, one might either attempt to find a single `best' choice of hyperparameter \(\boldsymbol{\theta}^*\), or one might consider a distribution of hyperparameters \(p_{\boldsymbol{\theta}}\) and consider an ensemble of fits \(\hat{f}_{\boldsymbol{\theta}}(x) = \mathbb{E}[f(x) | \boldsymbol{\theta},\mathbf{Y}]\) with different values of \(\boldsymbol{\theta}\).
In the first case, the hyperparameter might be chosen to optimize some objective function, such as a \textit{maximum likelihood estimator} (MLE)
\begin{equation*}
    \boldsymbol{\theta}_{MLE} := \operatorname{argmax}_{\boldsymbol{\theta}} p(\mathbf{Y}|\boldsymbol{\theta})
\end{equation*}
which tries to maximize the likelihood \(p(\mathbf{Y}|\boldsymbol{\theta})\) of observing the data \(\mathbf{Y}\) with prior GP specified by a given value \(\boldsymbol{\theta}\) of the hyperparameters.
\(p(\mathbf{Y}|\boldsymbol{\theta})\) is also known as the marginal likelihood, since it involves marginalizing over realizations \(f\), and usually the optimization is carried out over the log marginal likelihood for numerical reasons.

In the latter case, typically a \textit{hyperprior} \(q_{\boldsymbol{\theta}}\) is first specified, then Bayes' theorem is applied to find a posterior distribution which weights the hyperprior by the likelihood of observing the data
\begin{equation*}
    p_{\boldsymbol{\theta}}(\boldsymbol{\theta} | \mathbf{Y}) = q_{\boldsymbol{\theta}}(\boldsymbol{\theta}) \frac{p(\mathbf{Y}|\boldsymbol{\theta})}{\int p(\mathbf{Y}|\boldsymbol{\theta}) q(\boldsymbol{\theta}) d\boldsymbol{\theta}}
\end{equation*}
The inferred profile is then taken to be the average over the ensemble \(p_{\boldsymbol{\theta}}\) of profiles \(\hat{f}(x) = \mathbb{E}[\hat{f}_{\boldsymbol{\theta}}(x) | \mathbf{Y}]\), an approach known as \textit{(Bayesian) marginalization}.

Previous works have done comparisons between different methodologies for specifying \(\boldsymbol{\theta}\), in particular optimization versus marginalization \cite{chilenski_improved_2015}.
The difference between the uncertainty estimates produced by the two approaches can be understood through the law of total variance:
If \(\sigma^2_{\boldsymbol{\theta}}(x) := \operatorname{Var}[f(x) | \boldsymbol{\theta},\mathbf{Y}]\) is the uncertainty of an inference given a single point estimate of \(\boldsymbol{\theta}\), then the total uncertainty can be computed
\begin{equation} \label{eq:ltv}
    \operatorname{Var}[f(x) | \mathbf{Y}] = \mathbb{E}[\sigma^2_{\boldsymbol{\theta}}(x) | \mathbf{Y}] + \operatorname{Var}[\hat{f}_{\boldsymbol{\theta}}(x) | \mathbf{Y}]
\end{equation}
where the variance and expectation are taken over \(p_{\boldsymbol{\theta}}\).

The first term \(\mathbb{E}[\sigma^2_{\boldsymbol{\theta}}(x) | \mathbf{Y}]\) on the right hand side of \eqref{eq:ltv} is sometimes referred to as \textit{aleatoric} (within-model) uncertainty, and captures the averaged effect of how measurement uncertainties modeled by \(\boldsymbol{\varepsilon}\) propagate through to uncertainty in the inference for a given choice of kernel and associated hyperparameters \(\boldsymbol{\theta}\).
Note that GPR with a specified kernel is already a Bayesian method which posits an explicit hypothesis for the prior distribution of profiles \(f\) over which the inference is performed, hence it is appropriate to describe its output uncertainty as aleatoric.

The second term \(\operatorname{Var}[\hat{f}_{\boldsymbol{\theta}}(x) | \mathbf{Y}]\) is referred to as \textit{epistemic} (between-model) uncertainty, and captures uncertainty arising from variations in the ensemble of inferences due to varying choices of \(\boldsymbol{\theta}\).
If the variation of \(\boldsymbol{\theta}\) within the distribution \(p_{\boldsymbol{\theta}}\) produces variations in the inference \(\hat{f}_{\boldsymbol{\theta}}(x)\) that are larger than the typical uncertainty \(\sigma^2_{\boldsymbol{\theta}}(x)\) for fixed \(\boldsymbol{\theta}\), then the epistemic uncertainty will dominate the aleatoric uncertainty.

In the remainder of this work, we will primarily focus on GPR inferences for individual values of \(\boldsymbol{\theta}\).
These values might arise as the output of an optimization routine, or could represent individual members of an ensemble \(p_{\boldsymbol{\theta}}\).
When we present error bars for a fixed value of \(\boldsymbol{\theta}\), they should be understood as only representing the aleatoric component of the uncertainty.
Typically the marginalization approach will result in larger uncertainties due to the inclusion of epistemic uncertainty, especially for gradients.

As discussed earlier, profile fitting is severely undetermined for any finite number of measurements.
Typically, this problem must be addressed by including a degree of either implicit or explicit regularization during the profile inference process.
When adding regularization, two main issues must be avoided:
\begin{enumerate}
    \item (Over-regularization) The inference procedure should not inadvertently filter out structures or features present over the spatial or temporal scales of interest.
    \item (Over-fitting) The inference procedure needs to ensure that a sufficient diversity of measurements contribute to the inference of the profile at any given point, otherwise the fit may become overly sensitive to noise or systematic error in the measurements.
\end{enumerate}
Avoiding these issues is a necessary (although perhaps not sufficient) condition for the inferred profile to represent a credible interpretation of the measurements.

Our aim is to develop criteria for evaluating whether an inference for a given value of \(\boldsymbol{\theta}\) suffers from either over-regularization or over-fitting.
This involves translating the mathematical modeling choices made in GPR into an interpretable set of metrics which can be used to assess the credibility of an inference by comparing its degree of regularization with physically meaningful quantities.
Intuitively, different values of \(\boldsymbol{\theta}\) lead to a tradeoff between over-regularization and over-fitting.
It may be the case that there are no admissible values of the hyperparameter which can simultaneously meet the requirements to avoid over-regularization and over-fitting, in which case either more measurements or less demanding resolution requirements are needed.

In the context of optimization or point estimates of \(\boldsymbol{\theta}\), these metrics can inform the bounds on admissible values of the hyperparameter and determine if an individual inference can be taken as a credible (although perhaps non-exhaustive) interpretation of the data.
We will discuss these applications in more detail in the following sections.
Although we do not study marginalization in detail, these metrics can also be used to guide the design and interpretation of hyperpriors on \(\boldsymbol{\theta}\).
With a properly designed hyperprior, the fraction of hyperparameters in the posterior ensemble \(p_{\boldsymbol{\theta}}\) which meet the credibility criteria can be used to give an overall evaluation of the credibility of the inference.

\section{GPR as Low-Pass Filtering} \label{sec:lowpass}
In this section, we show how GPR can be interpreted as a form of low-pass filtering, and use this to give an explicit procedure for interpreting and choosing ranges of hyperparameters in the Mat\'{e}rn family of kernels, including the squared exponential kernel as a limit.

First, in section \ref{subsec:gpr_bias} we develop an analogy between GPR inference and low-pass filtering in an idealized infinite, regular-sampling scenario.
In particular, we give a simple expression for the cutoff frequency of the filter associated with the Mat\'{e}rn family.
We briefly discuss the multivariate case in section \ref{subsec:filter_multivariate}.

We follow this by providing specific recommendations suggested by this analysis in section \ref{subsec:lowpass_recomenndations}.
We emphasize the fact that the cutoff frequency, rather than depending on the length scale \(\ell\) alone, also depends on the quality of the data and the other hyperparameters through the signal-to-noise rate.
We also show how the usual squared exponential kernel can exhibit ringing artifacts due to band limiting of the associated low-pass filter, and demonstrate how Mat\'{e}rn kernels with \(\nu < \infty\) can alleviate these artifacts somewhat.

\subsection{Spectral Analysis of GPR Inference} \label{subsec:gpr_bias}

Suppose we have a bi-infinite sequence of measurements \(\{Y_{i}\}\) at regularly spaced points \(\{x_{i}\} = \{i \Delta x\}\) where \(i \in \mathbb{Z}\).
Furthermore, assume we take zero mean \(\mu = 0\), a stationary covariance kernel \(k(x,x') = \kappa(x-x')\) and diagonal noise \(\boldsymbol{\Sigma}_{\varepsilon} = \sigma_{\varepsilon}^2 I\).

In statistical inference problems, an estimator \(\hat{f}\) for a value \(f\) is called \textit{unbiased} if \(\mathbb{E}[\hat{f}|f] = f\), and \textit{biased} if \(\mathbb{E}[\hat{f}|f] \neq f\).
Notice that the conditional expectation is taken with respect to the value \(f\) we are trying to estimate, not with respect to the data \(\mathbf{Y}\).
Thus, we switch points of view and take the formula \eqref{eq:fhat} as the definition of \(\hat{f}(x)\), and try to see how \(\bar{f}(x) := \mathbb{E}[\hat{f}(x) | f]\) compares to \(f(x)\).
If \(\bar{f}(x) = f(x)\), then the estimator \(\hat{f}(x)\) will be unbiased.

Define the continuous Fourier transform for a function \(f\)
\begin{equation*}
    \mathcal{F}[f](\xi) = \int_{-\infty}^{\infty} f(x) e^{-\mathfrak{i}2\pi \xi x} \, \dd{x}
\end{equation*}
where \(\mathfrak{i}^2 = -1\) is the imaginary unit.
We work with real frequency \(\xi\) rather than angular frequency.

We focus on the case where \(\xi \Delta x \lesssim 1/2, \Delta x / \ell \ll 1\), and \(\mathcal{F}[f](\xi)\) decaying quickly enough so that aliasing effects from the finite sampling rate can be ignored.
Then in \ref{app:lowpass_derivation}, we show the estimator mean \(\bar{f}(x) = \mathbb{E}[\hat{f}(x)|f]\) satisfies
\begin{equation} \label{eq:ctsfilter}
    \mathcal{F}[\bar{f}](\xi) \approx \frac{\mathcal{F}[\kappa](\xi)}{\sigma_\varepsilon^2 \Delta x + \mathcal{F}[\kappa](\xi)} \mathcal{F}[f](\xi) =: H(\xi) \mathcal{F}[f](\xi)
\end{equation}
Here \(H(\xi)\) is the `transfer function' defined in terms of the the Fourier transform of the Mat\'{e}rn kernel, which in 1D is given by
\begin{equation} \label{eq:c_nu}
\begin{gathered}
    \mathcal{F}[\kappa](\xi) = \begin{cases}
        \sigma^2 \ell C_\nu \left(1 + \frac{(2 \pi \xi \ell)^2}{2\nu}\right)^{-\left(\nu+\tfrac{1}{2}\right)} & 0 < \nu < \infty \\
        \sigma^2 \ell \sqrt{2\pi} \exp(-\frac{(2 \pi \xi \ell)^2}{2}) & \nu = \infty
    \end{cases} \\
    C_\nu := \sqrt{2 \pi} \tfrac{\Gamma(\nu+\tfrac{1}{2})}{\Gamma(\nu) \nu^{1/2}}
\end{gathered}
\end{equation}
\(H(\xi)\) is also called the \textit{equivalent kernel} \cite{rasmussen_gaussian_2008}.

\begin{figure}
    \centering
    \includegraphics[width=\halfcolwidth]{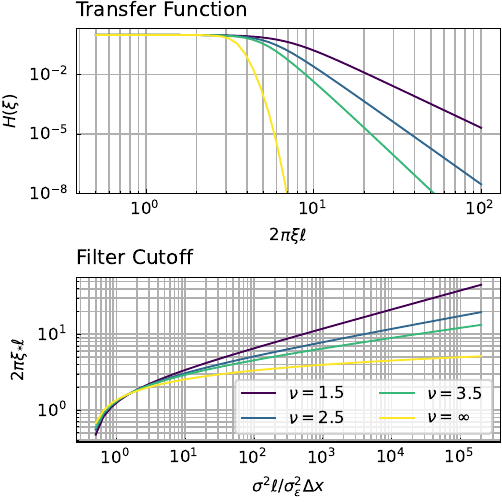}
    \caption{(Top) Mat\'{e}rn transfer functions \(H(\xi)\) against frequency \(\xi\) and (bottom) filter cutoffs \(\xi_*\) against the signal to noise rate \(S=\sigma^2 \ell / \sigma_\varepsilon \Delta x\) for different values of \(\nu\). Frequencies are normalized to \(2 \pi \ell\). GPR will heavily damp frequencies \(\xi\) above the cutoff frequency \(\xi_*\).}
    \label{fig:lowpass_halfwidth}
\end{figure}

Note that the right-hand side of \eqref{eq:ctsfilter} is in general not equal to \(\mathcal{F}[f](\xi)\).
Taking the inverse Fourier transform of both sides, we can conclude that \(\bar{f}(x) = \mathbb{E}[\hat{f}(x)|f]\neq f(x)\), so \(\hat{f}(x)\) is a biased estimator of \(f(x)\).
Plots of \(H(\xi)\) for different values of \(\nu\) are shown in figure \ref{fig:lowpass_halfwidth}.
When \(\mathcal{F}[\kappa] \gg \sigma_{\varepsilon}^2 \Delta x\), \(H(\xi) \approx 1\), which occurs for low frequencies.
Then, above some cutoff frequency \(\xi > \xi_*\), \(H(\xi) \to 0\).

This behavior for \(H(\xi)\) indicates that GPR inference \(\hat{f}(x)\) provides an estimate for a low-pass filtered version of \(f\), not \(f\) itself.
Low frequency components of \(f\) are inferred in an approximately unbiased way, but high frequency components of \(f\) are attenuated by the bias.
This phenomenon can be interpreted as a simple form of \textit{spectral bias}, observed in other machine learning settings \cite{rahaman_spectral_2019}.
An alternative interpretation of these results is that \(\hat{f}(x)\) is an unbiased estimator for the low-pass filtered version of \(f\).

We can compute the cutoff frequency \(\xi_*\) of the filter through the cutoff condition \(H(\xi_*) = 1/2\), which results in an explicit expression
\begin{equation} \label{eq:cutoff}
    \begin{gathered}
        2 \pi \xi_* \ell = \begin{cases}
            \sqrt{2 \nu} \left[\left(C_\nu S\right)^{\tfrac{2}{2 \nu + 1}} - 1\right]^{1/2} & 0 < \nu < \infty \\
            \left[2 \log(\sqrt{2\pi}S)\right]^{1/2} & \nu = \infty
        \end{cases} \\
        S := \sigma^2 \ell / \sigma_{\varepsilon}^2 \Delta x
    \end{gathered}
\end{equation}
where \(C_{\nu}\) is the same as in \eqref{eq:c_nu}.
\(S\) is the \textit{signal-to-noise rate}, which is the signal-to-noise ratio \(\sigma^2/\sigma_{\varepsilon}^2\), multiplied by the sampling rate normalized to the length scale of the kernel \(\ell / \Delta x\).
Plots of the filter cutoff as a function of the signal-to-noise rate for different values of \(\nu\) are shown in figure~\ref{fig:lowpass_halfwidth}.
We remark that \(H(\xi) \propto \xi^{-(2\nu + 1)}\) for \(\xi \gg \xi_*\), suggesting that \(2\nu + 1\) can be interpreted as the order of the low-pass filter.
We also note that a completely flat pass-band with exact cutoff would be achievable with a sinc kernel \cite{tobar_band-limited_2019}, which may be useful in some applications.

\subsection{Multivariate Case} \label{subsec:filter_multivariate}

It is possible to directly extend this analysis to the multivariate inference case.
Suppose the data are given on an \(n\)-dimensional lattice with spacing \(\Delta x_{(1)}, ..., \Delta x_{(n)}\) in each direction.
Denote the components of vectors by \(x=(x_{(1)}, ..., x_{(n)})\), and write \(\mathcal{F} = \mathcal{F}_{(1)}\circ ... \circ \mathcal{F}_{(n)}\) to be the Fourier transform in each of the components, which is equal to the composition of the Fourier transform in each of the individual components.
Assuming that aliasing effects can be ignored, equation \eqref{eq:nfilter} simplifies to
\begin{equation} \label{eq:nctsfilter}
    \mathcal{F}[\bar{f}](\xi) \approx \frac{\mathcal{F}[\kappa](\xi)}{\sigma_\varepsilon^2 |\Delta x| + \mathcal{F}[\kappa](\xi)} \mathcal{F}[f](\xi) =: H(\xi)\mathcal{F}[f](\xi)
\end{equation}
where \(|\Delta x| := \prod_{r=1}^{n}\Delta x_{(r)}\).

Notice in the multivariate inference case, \(H(\xi)\) is a function of the vector of frequencies \(\xi\).
The cutoff condition \(H(\xi_*)=1/2\) will generally define a co-dimension 1 hypersurface in frequency space enclosing the origin.
Frequencies inside this curve are part of the pass band, while frequencies outside of the curve will be attenuated by the filter.
We will defer discussion of how to characterize the cutoff frequencies of the multivariate case until the concrete example in section \ref{subsec:d3d_elm}.

\subsection{Recommendations from Low-Pass Interpretation} \label{subsec:lowpass_recomenndations}

\begin{figure*}
    \centering
    \includegraphics[width=\linewidth]{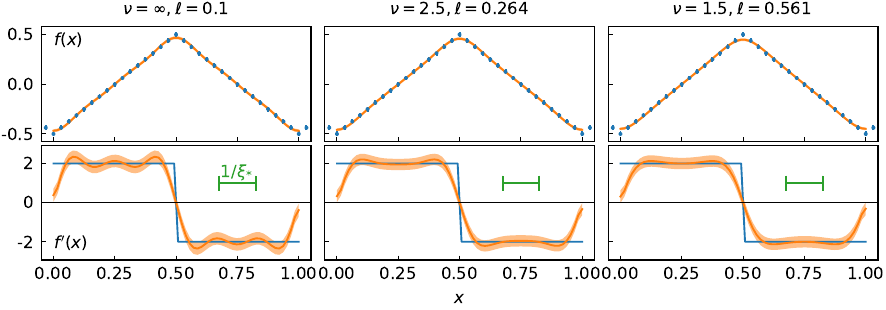}
    \caption{Illustration of `ringing' artifacts caused by the Gibbs phenomenon when using GPR to infer non-smooth functions. True profile (blue line) and data (blue scatter) are shown along with inferences (orange with \(\pm 34\%\) confidence interval shaded) overplotted. The inverse filter cutoff \(1/\xi_*\) is also shown.}
    \label{fig:lowpass_ringing}
\end{figure*}

One key observation from the preceding analysis is that the characteristic length scale of the inferred profile is not simply given by \(\ell\), and instead is determined by the cutoff frequency \(\xi_*\) which depends both on \(\ell\) and the other hyperparameters through the signal-to-noise rate \(S\).
Intuitively, since GPR doesn't constrain the prior functional form of the inference, functions with high-frequency components above the length scale \(\ell\) can be inferred if the quality of the data is sufficiently high.

Another key behavior of GPR revealed by this analysis is the possibility of `ringing' artifacts caused when the smoothness parameter \(\nu\) is large or infinite, but the function \(f(x)\) is not smooth.
Observe in figure \ref{fig:lowpass_ringing}, where we attempt to use GPR with three different values of \(\nu\) to fit samples from a triangle wave.
\(\sigma^2=1/12\) is picked to match the amplitude of the triangle wave and is the same between all three cases, while \(\ell\) is chosen such that the filter has the same cutoff frequency \(\xi_*\approx 6.6798\) for different values of \(\nu\).
Note that no noise has actually been applied to the synthetic data, despite the non-zero errorbars, in order to emphasize the ringing effect.
The \(\nu = \infty\) case shows ringing artifacts in the fit of the derivative \(\partial_x f\), which result from the Gibbs phenomenon when the high-frequency components are over-attenuated by the low-pass filter.
Decreasing \(\nu\) while maintaining the same filter cutoff frequency can help alleviate these ringing artifacts.

Knowledge of the low-pass filtering behavior of GPR can give some simple guidelines to inform the choice of hyperparameters:
\begin{enumerate}
    \item Variance \(\sigma^2\): A starting estimate for this parameter can be computed from the relation \(\mathbb{E}[(f(x)-\mu(x))^2] = \sigma^2\) by taking \(f(x_i) = Y_i\) as a first pass, then averaging \((Y_i - \mu(x_i))^2\) over the data points if they are roughly uniformly distributed over the domain.
    \item Smoothness \(\nu\): If the inferred function lacks sharp changes in any of its derivatives, \(\nu = \infty\) should be chosen since it has the sharpest drop-off in filtering high-frequency components.
    However, if a sharp change in any of the derivatives is expected, such as due to the presence of a localized transport barrier, finite \(\nu\) can be considered.
    For these problems, \(\nu = 2.5\) was found to strike a good balance between alleviating the Gibbs phenomenon while retaining a reasonable drop-off rate for high frequency noise.
    \item Length scale \(\ell\): For a given variance \(\sigma^2\) and smoothness \(\nu\), the cutoff frequency \(\xi_*\) can be computed in terms of \(\ell\) using \eqref{eq:cutoff}.
    Aliasing should usually be avoided, so the cutoff frequency must satisfy \(\xi_* \Delta x \lesssim 1/2\). The exact choice of \(\ell\) can then be chosen based on the resolution requirements through \(\xi_*\).
\end{enumerate}
To avoid over-regularization for pedestal profiles, it is reasonable to assert that profile features on the scale of the pedestal itself need be inferred.
This corresponds to a requirement that \(\xi_* w_{ped} \gtrsim 1\) for some measure \(w_{ped}\) of pedestal width.
The estimate of the pedestal width may itself depend on the degree of smoothing applied, so satisfying this requirement may involve a degree of exploration of the hyperparameter space, which we consider later in section \ref{sec:d3d}.

\begin{figure*}
    \centering
    \includegraphics[width=\linewidth]{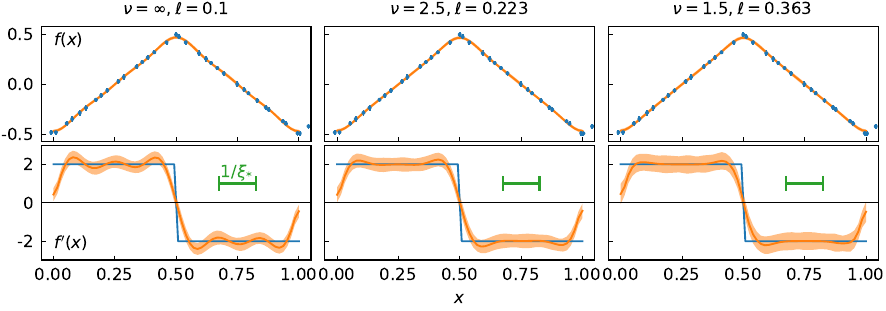}
    \caption{Illustration of effective low-pass filter behavior of GPR when applied to non-uniformly sampled data with non-identical error bars. Figure labels, as well as hyperparameters chosen for the GPR kernel, are the same as in Figure \ref{fig:lowpass_ringing}.}
    \label{fig:lowpass_ringing_sampled}
\end{figure*}

In practice, real experimental data is often given on irregularly spaced locations and with non-identical error bars.
One way to apply the low-pass filter analysis is to compute an average signal-to-noise rate
\begin{equation}
    \bar{S} = \sigma^2 \ell^2 / \overline{\sigma_\varepsilon^2} \overline{\Delta x}
\end{equation}
using the average spacing \(\overline{\Delta x}\) and average measurement variance \(\overline{\sigma_\varepsilon^2}\) for the data points in the region of interest.
A caveat for this approach is that the low-pass filter analogy and corresponding filter cutoff \(\xi_*\) will only give an approximation of the actual behavior of the GPR fit.
Figure~\ref{fig:lowpass_halfwidth} and equation \eqref{eq:cutoff} show that the scaling of \(\xi_*\) with \(S\) is either a fairly weak power law or logarithmic.
For \(S \gg 1\), which is the typical case, the dependence of \(\xi_*\) on \(\nu\) is much more impactful, so \eqref{eq:cutoff} may still give the correct overall trend of \(\xi_*\) using \(\bar{S}\), even in the case with irregularly spaced locations and non-identical error bars.

In figure \ref{fig:lowpass_ringing_sampled}, we demonstrate an example of a GPR inference which randomizes the measurement locations and error bars of the synthetic data from the example in figure~\ref{fig:lowpass_ringing}, while keeping the average signal-to-noise ratio fixed.
Again, note that no noise has been applied to the synthetic data to emphasize the effect of the low-pass filter.
The inferences do not deviate significantly from the regularly sampled identical measurement case shown in figure~\ref{fig:lowpass_ringing}, showing the relevance of the low-pass filter analogy when using an averaged signal-to-noise rate.

To close this section, we briefly discuss how this picture in terms of resolution requirements relates to the Bayesian approach involving hyperpriors.
The design of hyperpriors typically requires the incorporation of physical assumptions or intuition built across a particular class of profiles.
For a given hyperprior, evaluating the cutoff frequency across the range of posterior hyperparmeters gives a quantitative picture of how assumptions on the frequency content of \(f\) will bias the resulting estimate \(\hat{f}\).

If the computed cutoff frequencies do not meet the resolution requirements, one of three approaches can be taken to address issues of over-regularization depending on the circumstances.
The first is to attempt increase the signal-to-noise rate, in which case the formula for the cutoff frequency can give an order of magnitude estimate for the additional amount of data necessary to meet the resolution requirement.
If the set of measurements if fixed, the second approach is to see if the hyperprior can be redesigned to incorporate less restrictive hypotheses on the expected frequency content of the profiles, in which case the formula for the cutoff frequency gives a guide on how changes in the hyperparmeters will affect the resulting inference.
If neither approach works to address over-regularization, then the final approach is to conclude that the measurements cannot resolve the full desired range of frequencies.
Recalling that GPR gives an unbiased estimate for the low-pass filtered profile, the inferred profile can be presented along with the cutoff frequency as an estimate of the degree of over-regularization applied by the inference process.

\section{Information Theory for GPR} \label{sec:neff}

While in the previous section we developed a theory for GPR in an idealized case of filtering continuous functions, in this section we focus on information theory to answer the question: 
without making any simplifying assumptions, exactly how many measurements contribute to the inference of a value at a given point?

We begin by introducing the concept of Fisher information and its relation to inference problems in section \ref{subsec:fisherinfo}.
This is followed by the definition of \(N_{eff}\), a metric that measures the effective number of measurements which contribute to an inference, in section \ref{subsec:neff}.
Finally, we discuss some mathematical properties and examples of \(N_{eff}\) in section~\ref{subsec:neffprops}, and we discuss the relationship between \(N_{eff}\) and robustness of GPR inference to unmodeled errors in section~\ref{subsec:neffunmodeled}.

\subsection{Fisher Information} \label{subsec:fisherinfo}

As a simple motivating example, consider a sequence of independent, identically distributed random variables \(Y_i \sim \mathcal{N}(m,\sigma^2)\).
Suppose we wish to infer the mean \(m\) of these variables assuming \(\sigma^2\) is known.
If we define the sample mean
\begin{equation*}
	\hat{m}_N = \frac{1}{N} \sum_{i=1}^{N} Y_i
\end{equation*}
then this estimator satisfies
\begin{equation*}
	\mathbb{E}[\hat{m}_N] = m \qquad \mathbb{E}[(\hat{m}_N - m)^2] = \sigma^2/N
\end{equation*}
Notice the variance scales as \(\sim 1/N\), so the estimator should get better with more independent measurements as the usual \(\sim1/\sqrt{N}\) scaling.

To generalize this \(\sim 1/N\) scaling of the variance to other contexts, we consider it from an information-theoretic lens.
The Fisher information of a collection of real-valued continuous random variables \(\mathbf{Y}\) with respect to a set of \(M\) real-valued continuous parameters \(\boldsymbol{\theta}\) is defined as the \(M \times M\) matrix with components
\begin{equation}
    \left[\mathcal{I}_{\mathbf{Y}}[\boldsymbol{\theta}]\right]_{ij} := -\mathbb{E}\left[\frac{\partial^2}{\partial \theta_i \partial \theta_j}\log p(\mathbf{Y};\boldsymbol{\theta}) \middle| \boldsymbol{\theta} \right]
\end{equation}
where \(p(\mathbf{y};\boldsymbol{\theta})\) is the PDF of \(\mathbf{Y}\) for a given value of the parameter \(\boldsymbol{\theta}\) \cite{lehmann_theory_1998}.
Note the Fisher information is a positive semi-definite matrix.

One first key property of the Fisher information is that ``information is additive''.
In particular, if \(\mathbf{X},\mathbf{Y}\) are independent then
\begin{equation*}
    \mathcal{I}_{\mathbf{X},\mathbf{Y}}[\boldsymbol{\theta}] = \mathcal{I}_{\mathbf{X}}[\boldsymbol{\theta}] + \mathcal{I}_{\mathbf{Y}}[\boldsymbol{\theta}]
\end{equation*}
A second key property of the Fisher information is that ``more information gives a better estimate''.
That is, suppose \(\mathbf{T}(\mathbf{Y})\) is any \textit{unbiased estimator} of \(\boldsymbol{\theta}\), i.e. \(\mathbb{E}[\mathbf{T}(\mathbf{Y})] = \boldsymbol{\theta}\).
Then, under appropriate smoothness and decay conditions the \textit{Cram\'{e}r-Rao bound} holds:
\begin{equation*}
    \operatorname{Cov}[\mathbf{T}(\mathbf{Y})] \ge \mathcal{I}_{\mathbf{Y}}[\boldsymbol{\theta}]^{-1}
\end{equation*}
where \(\ge\) is in the sense of positive semi-definite matrices.

To interpret the second key property, note that the eigenvalues of the covariance matrix give a measure of the spread of the distribution of \(\mathbf{T}(\mathbf{Y})\), with smaller eigenvalues implying that the distribution of \(\mathbf{T}(\mathbf{Y})\) is more closely concentrated around the mean \(\boldsymbol{\theta}\).
One consequence of the matrix inequality above is that the smallest eigenvalue of \(\operatorname{Cov}[\mathbf{T}(\mathbf{Y})]\) is bounded below by the inverse of the largest eigenvalue of \(\mathcal{I}_{\mathbf{Y}}[\boldsymbol{\theta}]\), putting a limit on how tightly concentrated the distribution can be.

Combining these two properties allows for a generalization of the earlier example.
Taking \(Y_i\) and \(\hat{m}_N\) as before, one can show that the Fisher information satisfies \(\mathcal{I}_{Y}[m] = N \mathcal{I}_{Y_1}[m]\), i.e. the collection of \(N\) measurements contains \(N\) times more information than a single measurement.
Then, the Cram\'{e}r-Rao bound says
\begin{equation*}
    \operatorname{Var}[\hat{m}_N] \ge (N\mathcal{I}_{Y_1}[m])^{-1} \sim 1/N
\end{equation*}

Note that since \(\mathbf{T}(\mathbf{Y})\) is itself a random variable, it also makes sense to consider the Fisher information of \(\mathbf{T}(\mathbf{Y})\) with respect to \(\boldsymbol{\theta}\), which provides the tighter bound
\begin{equation*}
    \operatorname{Cov}[\mathbf{T}(\mathbf{Y})] \ge \mathcal{I}_{\mathbf{T}(\mathbf{Y})}[\boldsymbol{\theta}]^{-1}
\end{equation*}
The so-called \textit{data processing inequality} \cite{zamir_proof_1998} says that \(\mathcal{I}_{\mathbf{T}(\mathbf{Y})}[\boldsymbol{\theta}] \le \mathcal{I}_{\mathbf{Y}}[\boldsymbol{\theta}]\), which reduces this to the earlier expression.
When \(\mathcal{I}_{\mathbf{T}(\mathbf{Y})}[\boldsymbol{\theta}] = \mathcal{I}_{\mathbf{Y}}[\boldsymbol{\theta}]\) this tells us that our estimator is using all available Fisher information from \(\mathbf{Y}\) to inform our knowledge of \(\boldsymbol{\theta}\).
Note that the Fisher information, and more generally information geometry theory, has seen usage in plasma physics in both statistical modeling and inference \cite{kim_time-dependent_2020,kim_probabilistic_2025}.

\subsection{Definition of \(N_{eff}\)} \label{subsec:neff}

Now we return to the GPR inference problem as specified in section \ref{sec:setup}.
We focus on the case where the measurement errors \(\varepsilon_i \sim \mathcal{N}(0,\sigma_i^2)\) are distributed as independent Gaussian random variables with zero mean and variance \(\sigma_i^2\), although we remark that the analysis can be generalized to cases with correlated measurement error.
Notice that the inferred profile \eqref{eq:fhat} is an affine (i.e. shifted linear) combination of independent measurements \(\mathbf{Y}\), which makes it relatively straightforward to apply the tools of information theory to this case.

Taking again \(\bar{f}(x) = \mathbb{E}[\hat{f}(x) | f]\), we show in \ref{app:neff_derivation} that for each \(x\),
\begin{equation} \label{eq:info_sum}
    \mathcal{I}_{\hat{f}(x)}[\bar{f}(x)] = \mathcal{I}_{\mathbf{Y}}[\bar{f}(x)] = \sum_{i=1}^{N} \mathcal{I}_{Y_i}[\bar{f}(x)]
\end{equation}
\begin{equation} \label{eq:info_part}
    \mathcal{I}_{Y_i}[\bar{f}(x)] = \frac{\sigma_i^2 \beta_i(x)^2}{\left(\sum_{k=1}^{N} \sigma_k^2 \beta_k(x)^2\right)^2}
\end{equation}
In other words, the information that the estimator \(\hat{f}(x)\) provides about \(\bar{f}(x)\), which in section \ref{sec:lowpass} was interpreted as a low-pass filtered version of \(f(x)\), is equal to the total information \(\mathbf{Y}\) contains about \(\bar{f}(x)\).
Furthermore, this total information in turn can be decomposed as the sum \eqref{eq:info_sum} of the information \eqref{eq:info_part} that each measurement \(Y_i\) provides about \(\bar{f}(x)\).

Now, if each measurement contributed the same amount of information to \(\bar{f}(x)\), the number of measurements would be equal to \(N = \mathcal{I}_{\mathbf{Y}}[\bar{f}(x)] / \mathcal{I}_{Y_i}[\bar{f}(x)]\).
We can generalize this to the situation where the measurements contribute different amounts of information by dividing the total information by the weighted average information per measurement, which gives a weighted effective number of measurements which contribute to the total information:
\begin{equation}
    \bar{\mathcal{I}}_{\mathbf{Y}}[\bar{f}(x)] := \sum_{i=1}^{N} \mathcal{I}_{Y_i}[\bar{f}(x)] \left(\tfrac{\mathcal{I}_{Y_i}[\bar{f}(x)]}{\mathcal{I}_{\mathbf{Y}}[\bar{f}(x)]}\right)
\end{equation}
\begin{equation} \label{eq:neff}
    N_{\mathbf{Y}}[\bar{f}(x)] := \frac{\mathcal{I}_{\mathbf{Y}}[\bar{f}(x)]}{\bar{\mathcal{I}}_{\mathbf{Y}}[\bar{f}(x)]}
\end{equation}
Equation \eqref{eq:neff} represents the main result of this section.
\(N_{\mathbf{Y}}[\bar{f}(x)]\) measures the effective number of measurements which contribute to the inference of \(\bar{f}(x)\) at a given point \(x\).
Since the relative importance of different measurements to the GPR inference changes for different points \(x\), \(N_{\mathbf{Y}}[\bar{f}(x')]\) will typically vary as a function of \(x\).

We also define several variants of this formula,
As discussed in the introduction, we are often more interested in inferences of profile gradients, in which case one can consider the Fisher information of the measurements with respect to \(\bar{f}'(x)\) using the estimator \eqref{eq:fhat_deriv}, which satisfies
\begin{equation*}
    \mathcal{I}_{Y_i}[\bar{f}'(x)] = \frac{\sigma_i^2 \beta'_i(x)^2}{\left(\sum_{k=1}^{N} \sigma_k^2 \beta'_k(x)^2\right)^2}
\end{equation*}
and one can similarly define \(N_{\mathbf{Y}}[\bar{f}'(x)] = \mathcal{I}_{\mathbf{Y}}[\bar{f}'(x)] / \bar{\mathcal{I}}_{\mathbf{Y}}[\bar{f}'(x)]\).

Another common issue in practice is that we are not interested in how many individual measurements which contribute to an inference, but how many bins of data contribute to the inference.
For example, we might be interested in how many spatial channels or time slices an inference might depend on.
To notate this, let \(\mathbf{P} = P_1,..., P_M\) be a partition of \(\{1,...,N\}\), i.e. each \(P_j \subset \{1,...,N\}\) and \(\bigsqcup_{j=1}^{M} P_j = \{1,...,N\}\).
Then we can write
\begin{equation*}
    \mathcal{I}_{P_j}[\bar{f}(x)] = \sum_{i \in P_j} \mathcal{I}_{Y_i}[\bar{f}(x)]
\end{equation*}
\begin{equation*}
    \bar{\mathcal{I}}_{\mathbf{P}}[\bar{f}(x)] = \sum_{j=1}^{M} \mathcal{I}_{P_j}[\bar{f}(x)] \left(\tfrac{\mathcal{I}_{P_j}[\bar{f}(x)]}{\mathcal{I}_{\mathbf{Y}}[\bar{f}(x)]}\right)
\end{equation*}
\begin{equation} \label{eq:neff_bin}
    N_{\mathbf{P}}[\bar{f}(x)] := \frac{\mathcal{I}_{\mathbf{Y}}[\bar{f}(x)]}{\bar{\mathcal{I}}_{\mathbf{P}}[\bar{f}(x)]}
\end{equation}
We can of course also consider \(N_{\mathbf{P}}[\bar{f}'(x)]\) similarly.
This version of \(N_{eff}\) has the interpretation of being the effective number of bins of data which contribute to the inference at a given point \(x\).

For brevity we will write \(N_{eff} = N_{eff}(x) = N_{\mathbf{Y}}[\bar{f}(x)], N_{\mathbf{Y}}[\bar{f}'(x)], N_{\mathbf{P}}[\bar{f}(x)], N_{\mathbf{P}}[\bar{f}'(x)]\), with any possible binning and inferred value defined in text.

\subsection{Mathematical Properties and Simple Example of \(N_{eff}\)} \label{subsec:neffprops}

Defining a vector \(\mathbf{s}\) by components \(s_i = \mathcal{I}_{Y_i}[\bar{f}(x)]\), a direct calculation shows
\begin{equation}
    N_{\mathbf{Y}}[\bar{f}(x)] = \left(\frac{\norm{\mathbf{s}}_1}{\norm{\mathbf{s}}_2}\right)^2
\end{equation}
where \(\norm{\mathbf{s}}_p := (\sum_{i=1}^{N} |s_i|^p)^{1/p}\) is the \(\ell^p\) norm of \(\mathbf{s}\).
This leads to three key properties of \(N_{eff}\):
\begin{enumerate}
    \item (\textit{Boundedness}) If at least one \(s_i\) is non-zero, \(1 \le N_{eff} \le N\). Furthermore, \(N_{eff} = 1\) if and only if exactly one \(s_i\) is non-zero, and \(N_{eff} = N\) if and only if all of the \(s_i\) are non-zero and equal.
    \item (\textit{Subset property}) If \(R\) of the \(s_i\) are equal and the rest are zero, then \(N_{eff} = R\).
    \item (\textit{Scaling invariance}) If \(\mathbf{s}\) is multiplied by a constant, then \(N_{eff}\) remains unchanged.
\end{enumerate}
Property 1 follows from H\"{o}lder's inequality, while properties 2 and 3 follow from direct calculation.
Similar properties hold for the other variants of \(N_{eff}\).

Property 1 shows that \(N_{eff}\) respects the physical number of measurements, and cannot be below 1 or above \(N\).
Property 2 shows when \(R\) measurements contribute equally to an inference, \(N_{eff} = R\).
Property 3 is somewhat more subtle, and shows that \(N_{eff}\) gives a metric which is independent of the size of the error bars.
Roughly speaking, the Fisher information scales like the inverse of the variance, a result which is exact for Gaussian random variables.
The scaling invariance says that \(N_{eff}\) is insensitive to the absolute magnitude of the information, and is only sensitive to the relative contribution of the different measurements.
An inference with small error bars may still lack credibility if it only depends on a small number of measurements, as the magnitude of the error bars of the underlying data may be underestimated due to errors not captured in \(\varepsilon_i\).

We remark that property 3 also shows that \(N_{eff}\) is not a proxy for the total amount of information contributed by the measurements.
If that is desired, the total information \(\mathcal{I}_{\mathbf{Y}}[\bar{f}(x)]\) should be used directly instead.
The choice to normalize the total information by the weighted average information \(\bar{\mathcal{I}}_{\mathbf{Y}}[\bar{f}(x)]\) at the same location \(x\) leads to the scaling invariance, but other choices may lead to other metrics capturing other information-theoretic properties of the inference.

Figure \ref{fig:gpr_basis} shows a simple example of the \(N_{eff}\) metric in practice.
Here, we take \(N_{eff}(x) = N_{\mathbf{Y}}[\bar{f}(x)]\), which will capture  the number of measurements used to infer the low-pass filtered profile \(\bar{f}\) at a given point \(x\).
Notice for the case with smaller \(\ell\), the regression coefficient functions \(\beta_i(x)\) are sharply peaked near each data point, indicating that the inferred profile is dominated by the value of a single data point \(Y_i\) near these peaks.
This is reflected in \(N_{eff}(x)\), which at the \(x\) locations of the data points, \(N_{eff} \approx 1\) indicates that only one data point is being used to infer \(\bar{f}(x)\).
Between the data points, \(N_{eff} \approx 2\), which matches the intuition that \(\bar{f}(x)\) is an average of adjacent data points at those locations.

Meanwhile for the case with larger \(\ell\), the regression coefficients are broader, indicating that no single data point dominates the inference \(\bar{f}(x)\).
Inspection of the regression coefficient functions \(\beta_i(x)\) indicates that around \(4\) are non-zero at any given point \(x\).
Correspondingly, \(N_{eff} \approx 4\), indicating that effectively 4 data points contribute to the inferred value of \(\bar{f}(x)\) across most of the profile.
Notice that close to the boundaries \(x=0\) and \(x=1\), there are fewer data points and their corresponding regression coefficient functions become more peaked, indicating that fewer data points are contributing to the inference near the boundaries.
This behavior is also captured by \(N_{eff}\), which drops off near the boundaries.
In fact, notice that the vector components \(s_i \propto \beta_i(x)^2\), so \(N_{eff}\) can also be interpreted in terms of the weighted contribution of the different functions \(\beta_i(x)\) to \(\bar{f}(x)\).

\subsection{\(N_{eff}\) Controls Unmodeled Errors} \label{subsec:neffunmodeled}

We close with an important interpretation of \(N_{eff}\) involving the control of unmodeled errors, which are uncertainties in the data not captured by the noise covariance \(\Sigma_\varepsilon\).
Often, not all sources of error are explicitly or correctly accounted for in the model for the measurement noise \(\varepsilon_i\).
A common example of practical interest are systematic uncertainties in the measurements due to calibration errors.
In this section, we demonstrate how \(N_{eff}\) gives a rigorous bound to the degree to which a particular class unmodeled errors can propagate into GPR inferences, with larger \(N_{eff}\) guaranteeing that the unmodeled error will make a smaller impact.

We consider a simple model of the impact of unmodeled correlated error between different measurements.
As before, let \(\mathbf{P} = P_1,...,P_M\) be a partition of \(\{1,...,N\}\).
Suppose \(\boldsymbol{\delta} = [\delta_1,...,\delta_M]^T\) is a random vector of deviations.
We assume that each of the deviations \(\delta_j\) are independent from the other deviations and from the measurement errors \(\boldsymbol{\varepsilon}\).
We also assume the deviations have zero mean and finite variance, although we do not require the deviations to be normally distributed.

For each bin \(P_j\), we suppose that each measurement \(\tilde{Y}_i, i \in P_j\) differs from the idealized measurement \(Y_i\) by
\begin{equation*}
    \tilde{Y}_i = Y_i + \sigma_i \delta_j
\end{equation*}
In other words, for each bin \(P_j\), the measurement error for \(i \in P_j\) is equal to \(\tilde{\varepsilon}_i = \varepsilon_i + \sigma_i \delta_j\).
Using \(\tilde{\mathbf{Y}}\) in place of \(\mathbf{Y}\) in \eqref{eq:fhat} would produce a profile \(\tilde{f}(x)\) which differs from the idealized inference \(\hat{f}(x)\) by
\begin{equation*}
    \tilde{f}(x) - \hat{f}(x) = \sum_{j=1}^{M} \left[ \delta_j \sum_{i\in P_j} \beta_i(x) \sigma_i \right]
\end{equation*}

Define the normalized deviation \(\tilde{\Delta}(x)\) by
\begin{equation}
    \tilde{\Delta}(x) := \frac{\operatorname{Var}[\tilde{f}(x) - \hat{f}(x) | f]}{\bar{\sigma}^2_\varepsilon(x)}
\end{equation}
where \(\bar{\sigma}^2_\varepsilon(x)\) is the weighted measurement noise
\begin{equation*}
    \bar{\sigma}^2_\varepsilon(x) := \sum_{i=1}^{N} \beta^2_i(x) \sigma_i^2 = \operatorname{Var}[\hat{f}(x) | f]
\end{equation*}
\(\tilde{\Delta}\) quantifies the mean squared deviation between the inferred profile \(\tilde{f}\) with an idealized inference \(\hat{f}\) which has the deviation \(\boldsymbol{\delta}\) subtracted out.
From a frequentist view, we can imagine this variance as computed over an ensemble of experiments which measure an identical plasma profile \(f\), where each experiment has a different realization of the noise \(\boldsymbol{\varepsilon}\) and systematic deviation \(\boldsymbol{\delta}\).
Note that similar definitions can be given for the derivative by replacing \(\beta_i(x)\) by \(\beta_i'(x)\).

Then, this mean squared deviation satisfies
\begin{equation} \label{eq:neff_inequality}
    \tilde{\Delta}(x)^2 \le \frac{\overline{D^2}}{N_{\mathbf{P}}[\bar{f}(x)]}
\end{equation}
where \(\overline{D^2} := \sum_{j=1}^{M} |P_j|^2 \operatorname{Var}[\delta_j]^2\) is the sum of the squared variances weighted by the number of elements \(|P_j|\) in each bin.
That is, given a fixed magnitude of the deviations as quantified by \(\overline{D^2}\), the resulting deviation in the inference \(\tilde{f} - \hat{f}\) can be reduced by increasing \(N_{eff}\).
Note that this result does not hold for systematic deviations which are correlated between different bins; in that case, a different binning must be used.

The proof of \eqref{eq:neff_inequality} is a straightforward application of the Cauchy-Schwarz inequality, which we show in \ref{app:neff_inequality}.
We also discuss an analogous result for the derivative, and give a simple example that achieves the worst case predicted by the bound in the appendix.
We remark that the origin of the squared variance scaling comes from the choice to weight the average information per measurement by the information itself which leads to \(\bar{\mathcal{I}}_{\mathbf{P}} \sim \mathcal{I}_{P_j}^2\).
Different choices of weighting could potentially lead to other information-theoretic inequalities which control the unmodeled error.
Other choices for capturing the unmodeled error may also suggest different information-theoretic metrics to consider.
We leave systematic study of these choices and their empirical behavior to future work.

\section{Application to DIII-D Pedestal Data} \label{sec:d3d}

In this section, we focus on applying the developed techniques to fit pedestal profiles from the DIII-D tokamak \cite{luxon_design_2002}, discharge \#174864.
This discharge, which comes from a series of discharges studying the physics of microtearing modes (MTMs) in DIII-D \cite{nelson_time-dependent_2021}, has a particularly wide pedestal as well as sufficient time resolution from the measurements to provide some information about the dynamics of the pedestal between edge localized modes (ELM).
In section \ref{subsec:d3d_avg} we consider profiles depending on the radial coordinate only, while in sections \ref{subsec:d3d_elm} and \ref{subsec:d3d_elm_results} we see how the tools can be used to verify the credibility of features in time-dependent profiles.

\subsection{ELM-averaged Fits} \label{subsec:d3d_avg}

\begin{figure*}
    \centering
    \includegraphics[width=1.0\linewidth]{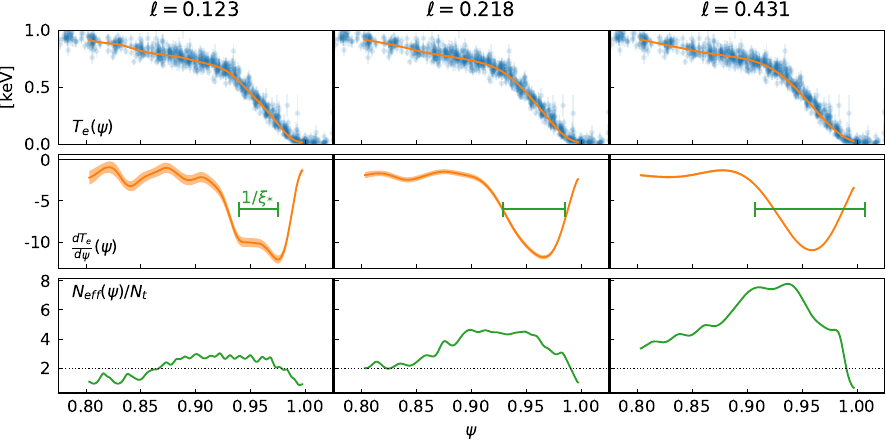}
    \caption{GPR inferences for the electron temperature \(T_e\) as a function of normalized poloidal flux \(\psi\) for DIII-D shot \#174864. (Top and middle rows) Data (blue scatter) are shown with corresponding inferred profiles (orange with \(\pm 34\%\) confidence interval shaded). The inverse filter cutoff \(1/\xi_*\) is also shown. Comparing to the width of the pedestal, the right-most profile shows signs of over-regularization. (Bottom row) \(N_{eff}\) for the derivative profiles is shown as a function of \(\psi\), with \(N_{eff} \lesssim 2\) indicating overfitting.}
    \label{fig:pedestal}
\end{figure*}

We begin by considering the problem of inferring the profile of the electron temperature \(T_e\) as a function of the normalized poloidal flux \(\psi\) from measurements from the Thomson scattering system \cite{eldon_initial_2012}.
Data are aggregated from \(N_t = 44\) time slices of Thomson data taken over 200ms, which cover ELM phases from 5\% to 90\% over two ELM cycles observed from 2.8s to 3s in the discharge.
The data, as well as GPR inferences for three different values of hyperparameters, are shown in figure \ref{fig:pedestal}.

The leftmost column displays an inference based on hyperparameters \(\sigma^2,\ell\) set to their MLE values while keeping \(\nu = 2.5\) fixed.
Some `wiggles' appear in the inferred derivative profile, suggesting the possibility of over-fitting.
To quantify this, we consider \(N_{eff}(\psi) = N_{\mathbf{Y}}[T_e'(\psi)]\), which represents the effective number of measurements contributing to the inference of the profile derivative \(T_e'\) at a given radial location \(\psi\).
The bottom row of figure \ref{fig:pedestal} plots this quantity normalized by the number of time slices \(N_t\).
The resulting normalized metric \(N_{eff}(\psi)/N_t\) provides the effective number of measurements per time slice used to infer the derivative at \(\psi\).
\(N_{eff}/N_t \lesssim 2\) indicates  that the derivative is primarily inferred using `sub-channel' resolution arising from the radial scatter of the measurements over different time slices.
This radial scatter introduces potential covariances in the data that may not be fully captured by the assumed model for the measurement error \(\varepsilon_i\), suggesting the MLE inference could be over-fitting fluctuations in the data in these regions.

\begin{figure}
    \centering
    \includegraphics[width=\halfcolwidth]{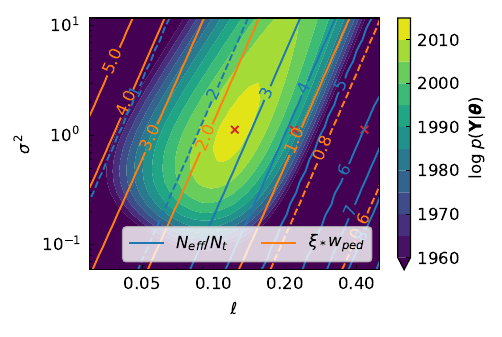}
    \caption{Filled contour plot of the log likelihood \(\operatorname{log} p(\mathbf{Y} | \boldsymbol{\theta})\) for various values of \(\ell,\sigma^2\). The MLE estimate, along with the cases with increased \(\ell\), are shown in red crosses. Overplotted on the likelihood contours are contour lines showing the median value of \(N_{eff}/N_t\) over the strong gradient region, as well as the low-pass filter frequency multiplied by the pedestal width \(\xi_* w_{ped}\). Dashed contour lines indicate inadmissible hyperparameter values either due to over-fitting or over-regularization.}
    \label{fig:pedestal_lml}
\end{figure}

\begin{figure}
    \centering
    \includegraphics[width=\halfcolwidth]{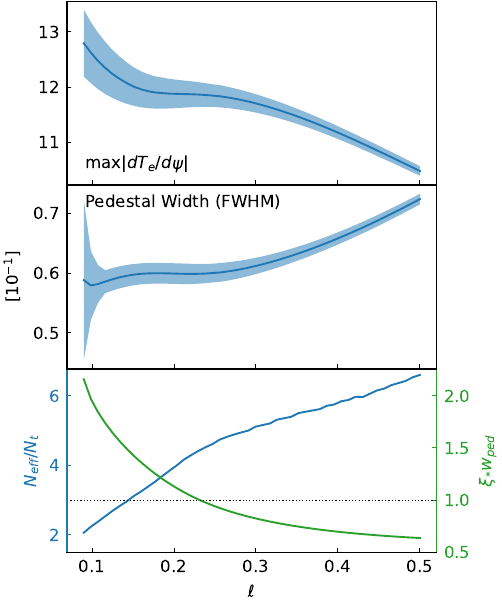}
    \caption{The top two plots show how varying the length scale \(\ell\) hyperparameter changes the inferred values for the peak gradient \(\operatorname{max}|dT_e/d\psi|\) and full width at half maximum (FWHM) \(w_{ped}\) of the temperature pedestal. The bottom plot shows how the median \(N_{eff}/N_t\) (blue) and the cutoff frequency normalized to pedestal width \(\xi_* w_{ped}\) (green) change as \(\ell\) is varied. The dashed line marks \(N_{eff}/N_t=3\) and \(\xi_* w_{ped} = 1\).}
    \label{fig:pedestal_scan}
\end{figure}

To explore the hyperparameter space relative to the MLE value, we plot the log likelihood \(\log p(\mathbf{Y} | \boldsymbol{\theta})\) in figure~\ref{fig:pedestal_lml}.
Overplotted in blue are contours of \(N_{eff}/N_t\), which show that as \(\ell\) is increased, the risk of over-fitting decreases.
Exploring other possible hyperparameter values, the cetner and rightmost columns of figure~\ref{fig:pedestal} show inferences with increased \(\ell\) relative to the MLE case.
These values could plausibly arise from a weighted objective function or hyperprior which biases the hyperparmeter towards larger length scale parameters.

For both of these inferences, \(N_{eff}(\psi)/N_t \gtrsim 4\) in the steep gradient region, indicating that the inferred derivative values are using information over multiple measurement channels.
This reduces concerns about overfitting sub-channel fluctuations in the data associated with the radial scatter.
However, there are significant quantitative differences in the inferred profiles for these two cases.
The rightmost case appears much smoother, and infers peak gradients that are 8\% smaller along with a pedestal width \(w_{ped}\), as measured by the full width at half maximum (FWHM) of the gradient profile, that is 13\% larger compared to the center case.

In order to systematically quantify the level of regularization applied by GPR, we turn to the low-pass analogy.
We consider an effective \(\overline{\Delta \psi} = 0.2 / N_{[0.8,1.0]}\), where \(N_{[0.8,1.0]}\) is the total number of measurements of the interval \(0.8 \le \psi \le 1.0\), and \(\overline{\sigma_\varepsilon^2}\) given by the average squared error of the data.
We can then compute an effective cutoff frequency \(\xi_*\) by using \eqref{eq:cutoff} with the average signal-to-noise rate \(\bar{S}=\sigma^2\ell/\overline{\sigma_{\varepsilon}^2}\overline{\Delta \psi}\).
We plot \(\xi_* w_{ped}\) in figure \ref{fig:pedestal_lml}, showing that the centermost inference satisfies \(\xi_* w_{ped} \gtrsim 1\), whereas the right-most inference does not.

In addition to plotting the inverse cutoff frequency \(1/\xi_*\) in figure~\ref{fig:pedestal}, we also show how the peak gradients, pedestal FWHM, \(N_{eff}/N_t\), and \(\xi_* w_{ped}\) systematically change as we vary \(\ell\) in figure \ref{fig:pedestal_scan}.
When \(\xi_* w_{ped}\) becomes less than unity, the inferred peak gradient decreases and pedestal width increases.
This suggests that physical frequencies associated with the pedestal structure are being filtered out by the inference for large values of \(\ell\), indicating over-regularization.
Meanwhile, when \(N_{eff}/N_t\) approaches 2, both the inferred peak gradient and the uncertainty in the pedestal width increase.
This reflects the inclusion of noise-sensitive higher frequencies for small values of \(\ell\), indicating possible over-fitting.
Taken together, the \(N_{eff}\) metric and the low-pass filter interpretation lend a high degree of confidence in the credibility of the inference in the center column of figure \ref{fig:pedestal} given this point estimate of the hyperparmeters \(\ell\).
\(N_{eff}/N_t \gtrsim 4\) in the peak gradient region provides confidence that the pedestal profile is not over-fit, while the low-pass cutoff frequency \(\xi_* w_{ped} \gtrsim 1\) provides confidence that the pedestal profile is not over-regularized.

Since the center column inference relies on a point estimate of the hyperparameters, it is natural to ask whether it provides an adequate representation of uncertainty compared to a full Bayesian marginalization.
Referring to the law of total variance \eqref{eq:ltv}, the variation in the inferred quantities as \(\ell\) varies in figure~\ref{fig:pedestal_scan} gives a simple estimate of how the error bars would change using marginalization.
For \(0.123 \le \ell \le 0.217\), these variations remain within the error bars computed from point estimates, indicating that epistemic uncertainty will not dominate over aleatoric uncertainty in this range.

However, it is important to emphasize that the inference \(\hat{f}\) represents an approximately low-pass filtered version of \(f\).
Therefore, it cannot be used to draw conclusions about the presence or absence of features at higher frequencies.
The \(N_{eff}\) metric and filter cutoff indicate that the available measurements are only sufficient to discriminate pedestal structure within a narrow range of scales.
For \(\ell > 0.217\) the hypotheses are too restrictive on the typical scales for \(f\), resulting in over-regularization that smooths away pedestal structure.
For \(\ell < 0.123\) the hypotheses are overly permissive and risk over-fitting.
The small epistemic uncertainty thus reflects the limited range of admissible hypotheses on the frequency content of \(f\) which are both consistent with the spatial resolution requirements to resolve pedestal structure, and robust against over-fitting.
Reminiscent of the `coastline paradox', the peak gradients could in principle become arbitrarily large as higher frequency fluctuations are included, but such possibilities are explicitly filtered out by the inference.
With these caveats in mind, and recognizing that unmodeled systematic errors can also increase the size of the error bars, the inference in the center column of figure~\ref{fig:pedestal} can be taken as a credible interpretation of the experimental measurements that provides an estimate of the low-pass filtered version of the pedestal profile.

\subsection{Inter-ELM Variation: Setup} \label{subsec:d3d_elm}

Now, we consider a powerful feature of GPR, which is its ability to fit continuously varying profiles of both time and space without needing to provide explicit parameterizations for the time variation.
Specifically, we look at the problem of inter-ELM evolution of pedestal electron pressure profiles, which is the period after an ELM crash where pedestal density and temperature profiles are re-established.
We use GPR to perform \textit{conditional averaging} of the ELM profiles, which is a method for aggregating measurement data from multiple ELM cycles to generate a single ensemble-averaged temporal history for the inter-ELM profile evolution.
This method works well when the ELMs behave in a regular, periodic manner.

To accomplish this, we aggregate measurements over \(N_t = 98\) time slices covering 5 ELM cycles from 2.8s to 3.7s in the discharge.
Data from ELMs with irregular frequency during this interval were discarded.
We assume that the value of a measurement \(Y_i\) depends on both the spatial and temporal location of its measurement, \(x_i=(\psi_i,t_i)\).
To perform the conditional average, we take \(t_i\) to be time since last ELM of the measurement.
Thus, we are trying to infer a two-dimensional function \(p(\psi,t)\) from noisy measurements \(Y_i = p(\psi_i,t_i) + \varepsilon_i\).
We take a GP prior on \(p\) with zero mean, and a covariance which is the direct product of two kernels,
\begin{equation} \label{eq:elm_kernel}
    \operatorname{Cov}[p(\psi,t)p(\psi',t')] = \sigma^2 \kappa_{\psi}(|\psi-\psi'|)\kappa_{t}(|t-t'|)
\end{equation}
We take \(\kappa_{\psi}, \kappa_{t}\) to be Mat\'{e}rn kernels associated with the spatial and temporal evolution, respectively.

Note that all of the dimensionality associated with the problem is captured by the functional dependence of the kernel; once the covariance matrix \(k(\mathbf{x},\mathbf{x})\) has been computed, the formula \eqref{eq:fhat} holds exactly as before.
The variance parameters \(\sigma^2_\psi,\sigma^2_{t}\) associated with the spatial and temporal portions of the kernel can be absorbed into a single hyperparameter \(\sigma^2\), so there are a total of five hyperparameters: \(\sigma^2,\nu_{\psi}, \ell_{\psi},\nu_{t}, \ell_{t}\).

We now use the tools developed in the previous sections to get a point estimate for a set of hyperparameters which provide a credible inference of the inter-ELM evolution.
Again, we emphasize that the same caveats discussed before still apply here.
Credibility should be understood in terms of giving a point estimate of a low-pass filtered version of the inter-ELM evolution meeting the resolution requirements.
Moreover, any error bars need to be understood as representing only the aleatoric uncertainty of this point estimate.

\begin{figure}
    \centering
    \includegraphics[width=\halfcolwidth]{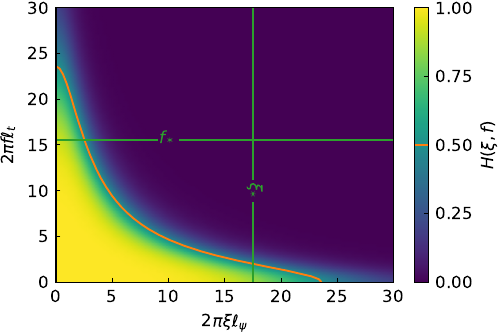}
    \caption{Plot of the two-dimensional transfer function \(H(\xi,f)\) over spatial and temporal frequencies \(\xi\) and \(f\) respectively. The filter cutoff curve \(H(\xi,f)=1/2\) is shown (orange curve), along with the effective cutoff frequences \(\xi_*,f_*\) (green lines) used to design the filter.}
    \label{fig:elm_transferfunc}
\end{figure}

\begin{figure}
    \centering
    \includegraphics[width=\halfcolwidth]{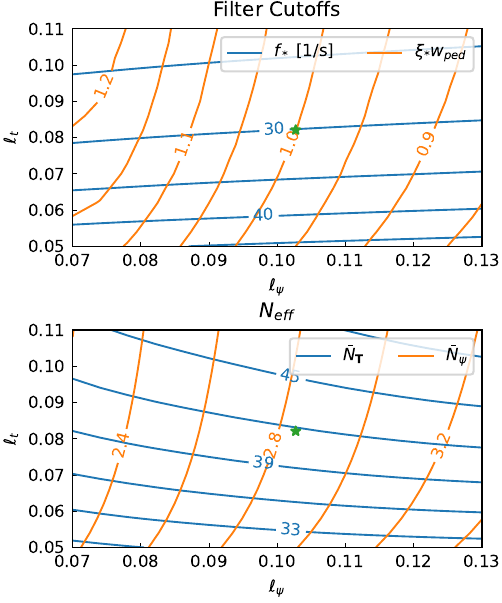}
    \caption{Impact of the length \(\ell_\psi\) and temporal \(\ell_t\) scale hyperparameters on (top) filter cutoffs and (bottom) \(N_{eff}\). For the filter cutoffs, the temporal filter cutoff \(f_*\) is shown (blue curves), along with the spatial filter cutoff times the pedestal width \(\xi_* w_{ped}\) (orange curves). The intersection of \(f_*=30\) and \(\xi_* w_{ped} = 1\) (green star) gives a candidate choice for \(\ell_\psi,\ell_t\). Similarly for the \(N_{eff}\) plot, the effective number of temporal bins used per inference \(N_{\mathbf{T}}\) (blue curves) are plotted along with the effective number of spatial points per time bin \(N_\psi\) (orange curves). \(f_*\) is given in units of cycles per second, while all other plotted quantities are dimensionless.}
    \label{fig:elm_resolution}
\end{figure}

Since we are fitting a pedestal profile, we use \(\nu_{\psi} = 2.5\).
Also, to capture sharp transitions in temporal behavior, we use \(\nu_t = 2.5\).
We get an estimate for \(\sigma^2\) by taking the mean square of the data \(Y_i^2\).
This leaves the two length scale parameters \(\ell_{\psi}, \ell_{t}\) to be determined.
In figure~\ref{fig:elm_transferfunc}, we use equation \eqref{eq:nctsfilter} to plot the transfer function \(H(\xi,f)\) associated with the product kernel \eqref{eq:elm_kernel} over spatial \(\xi\) and temporal \(f\) frequencies for a representative choice of \(\ell_\psi\) and \(\ell_t\).
Notice that the shape of the pass band enclosed by the \(H(\xi,f)=1/2\) contour indicates that high temporal frequencies experience stronger attenuation in spatial frequencies, and vice versa.

While knowledge of the entire transfer function is useful for detailed analysis of a fixed value of the hyperparameters, in order to use this information to inform our choice for \(\ell_\psi,\ell_t\) we need to compute summary values for the spatial and temporal cutoff frequencies.
While many different summary values exist, we found that a particularly good choice is to summarize the cutoffs with \textit{information-theoretic effective cutoffs} \(\xi_*,f_*\) defined as follows:

Let \(\mathbf{T}\) be the partition of the data into time slices, so two measurements \(Y_i,Y_j\) with \(i,j \in T_k\) are in the same bin \(T_k \in \mathbf{T}\) if \(t_i=t_j\).
Then, \(N_{\mathbf{T}}[\partial_\psi \bar{p}(\psi,t)]\) will measure the effective number of time slices from which data is used to infer the smoothed derivative \(\partial_\psi \bar{p}\) at a given \(\psi,t\).
Next, we need a metric to measure the effective number of measurement channels per time slice used to infer the derivative.
Due to the radial scatter of the points, for this particular case it does not make sense to directly bin the measurements into radial channels.
Instead, we define
\begin{equation*}
    N_\psi[\partial_\psi \bar{p}(\psi,t)] := \frac{N_{\mathbf{Y}}[\partial_\psi \bar{p}(\psi,t)]}{N_{\mathbf{T}}[\partial_\psi \bar{p}(\psi,t)]}
\end{equation*}
\(N_\psi\) is the ratio of the total number of measurements across all time slices used to infer \(\partial_\psi \bar{p}\) at a given \(\psi,t\) divided by the effective number of time slices used to infer the measurement.
Hence, it captures the effective number of measurements per time slice used for the inference.

Next, we average the values of \(N_{eff}\) over the spatial range \([0.8,1.0]\) and temporal range \([t_{min},t_{max}]\) of the data to compute average values \(\bar{N}_{\mathbf{T}}\) and \(\bar{N}_\psi\).
These are used to define the effective spatial and temporal signal-to-noise rates
\begin{equation*}
    \begin{gathered}
        \bar{S}_\psi := \frac{\sigma^2 \ell_\psi}{\overline{\sigma_\varepsilon^2} \overline{\Delta\psi}} \bar{N}_{\mathbf{T}} \\
        \bar{S}_t := \frac{\sigma^2 \ell_\psi}{\overline{\sigma_\varepsilon^2} \overline{\Delta t}} \bar{N}_{\psi}
    \end{gathered}
\end{equation*}
where \(\overline{\Delta \psi}\) is the average radial spacing between measurements from a single timeslice, while \(\overline{\Delta t}\) is the average temporal spacing between timeslices.
\(\bar{S}_\psi\) is the signal-to-noise rate for a spatial inference involving a single timeslice of data, \(\sigma^2 \ell_\psi / \overline{\sigma_\varepsilon^2} \overline{\Delta \psi}\), multiplied by the effective number of timeslices used in the inference of the profile derivative.
Similarly, \(\bar{S}_t\) is the signal-to-noise rate for a temporal inference involving a single measurement channel multiplied by the effective number of spatial channels used per timeslice in the inference of the profile derivative.
Plugging in \(S=\bar{S}_\psi\) into \eqref{eq:cutoff} gives a formula for an effective spatial cutoff frequency \(\xi_*\), and similarly \(S=\bar{S}_t\) gives an effective temporal cutoff frequency \(f_*\).
These effective cutoffs are also plotted for the representative case in figure~\ref{fig:elm_transferfunc}.

By keeping the hyperparameters \(\sigma^2,\nu_\psi,\nu_t\) fixed, we can plot \(\bar{N}_{\mathbf{T}}, \bar{N}_\psi,\xi_*\) and \(f_*\) as functions of \(\ell_\psi, \ell_t\).
This is done in figure~\ref{fig:elm_resolution}.
The ELM period was about \(\sim 100\) ms in the considered time range, so a temporal cutoff frequency \(f_* = 30\) cycles per second was chosen, so phenomena with a timescale of \(1/f_* \approx 33\) ms could be resolved.
We remark that without a clear understanding of the physical mechanisms controlling the inter-ELM temporal evolution, it is not possible to conclude if this timescale is sufficient to resolve all of the relevant physical timescales.
However, since \(f_*\) gives an estimate of the temporal smoothing applied by GPR, it is possible to conclude that the inference does not over-regularize on the macroscopic \(\sim 100\) ms ELM period timescale.
The spatial frequency \(\xi_*\) was chosen so that the product with the pedestal FWHM \(w_{ped}\) would satisfy \(\xi_* w_{ped} = 1\).

The intersection of the \(f_*=30\) and \(\xi_* w_{ped}=1\) contours then singles out a candidate choice of \(\ell_\psi,\ell_t\) which does not `over-regularize' the fit.
The plot of \(N_{eff}\) on the bottom of figure~\ref{fig:elm_resolution} provides confidence that this fit is not `over-fitting' either.
\(\bar{N}_{\mathbf{T}} \approx 42\) indicates that on average, effectively 42 time slices worth of data are used to infer the derivative at any given \(\psi, t\).
Meanwhile, \(N_\psi \approx 3\) indicates that 3 measurement channels per timeslice are used to infer the derivative on average.
We proceed with this point estimate of hyperparameters to analyze the information that time-dependent GPR can provide.

\subsection{Inter-ELM Variation: Results} \label{subsec:d3d_elm_results}

\begin{figure}
    \centering
    \includegraphics[width=\halfcolwidth]{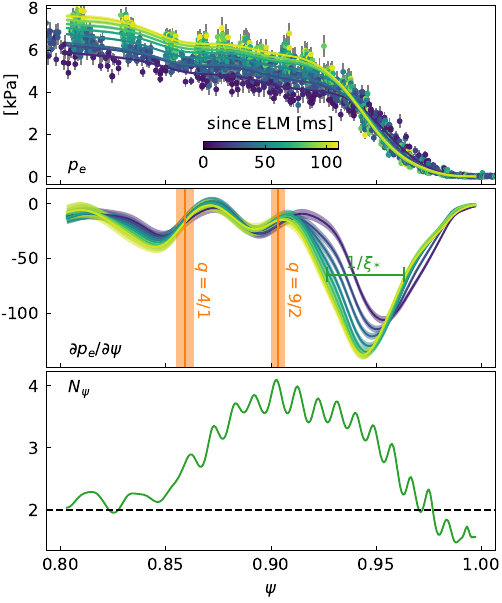}
    \caption{Time-dependent GPR inferences performed using the conditional averaging technique for the electron pressure profile \(p_e = n_e T_e\) as a function of \(\psi\). (Top and middle rows) Data (scatter) are shown with inferred profile and derivatives \(dp_e/d\psi\) (curves with \(\pm 34\%\) confidence intervals shaded), colored by the time since ELM. The inverse filter cutoff \(1/\xi_*\) is shown along with low-order rational \(q\) surfaces on the derivative profile. (Bottom row) \(N_{\psi}\), averaged over \(t\), is plotted as a function of \(\psi\). \(N_{\psi} \lesssim 2\) indicates possible overfitting due to reliance on two or less measurement channels per timeslice for the inference of the derivative.}
    \label{fig:elm_fit}
\end{figure}

\begin{figure}
    \centering
    \includegraphics[width=\halfcolwidth]{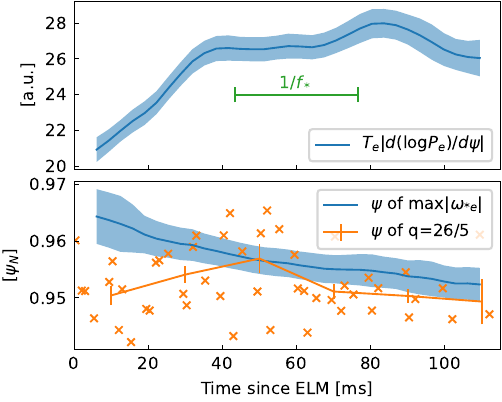}
    \caption{Evolution of summary statistics of from the time-dependent GPR inferences as a function of time since ELM. (Top) The peak of \(T_e |d(\log p_e)/d\psi | \propto \omega_{*,e}\) in the pedestal, proportional to the peak electron diamagnetic frequency, is shown along with the inverse temporal filter cutoff \(1/f_*\) (green bar).
    (Bottom) The location of the peak of \(\omega_*\) is shown (blue curve), along with the location of the \(q=26/5\) surface (orange line and scatter) for comparison.}
    \label{fig:elm_evolution}
\end{figure}

Recent work has shown how the alignment between the peak of the profile of electron diamagnetic frequency \(\omega_*\) with low-order rational surfaces in the pedestal can be used to predict features in experimentally observed magnetic fluctuation spectrograms \cite{hatch_microtearing_2021,larakers_global_2021,hassan_identifying_2022,curie_survey_2022,curie_gyrokinetic_2022}.
Computing profile gradients is crucial to such an analysis, as \(\omega_*\) is related to pressure gradients via
\begin{equation}
    \omega_* = \frac{\Phi_{sep}}{\psi_{sep}} n \frac{2 T_e}{a^2 e B_0} \frac{\partial(\log p_e)}{\partial\psi}
\end{equation}
where \(\Phi_{sep},\psi_{sep}\) are the toroidal and poloidal flux values at the separatrix, \(n\) is the toroidal mode number, and \(e B_0\) is the electron charge times the reference magnetic field, and \(\psi\) is the normalized poloidal flux.

In figure~\ref{fig:elm_fit}, we show the results of the time-dependent GPR inferences for the time-dependent electron pressure profile \(p_e = n_e T_e\) over an ELM cycle using the hyperparameters identified in the previous section.
The pressure measurements were derived by multiplying measurements of the electron density and temperature, with the error bars computed by propagation of error \(\sigma_p^2 = \sigma_n^2 T_e^2 + n_e^2 \sigma_T^2\).

Furthermore, following the analysis \cite{nelson_time-dependent_2021}, we are able to compare these pressure profiles against the locations of low-order rational \(q\) surfaces.
Namely, kinetic equilibria are produced every \(20\) ms using the CAKE code \cite{xing_cake_2021}, then sorted by the time since ELM.
The locations of low-order rational \(q\) surfaces are computed by averaging over their locations determined from the individual equilibria, and errorbars are computed using the standard error of the mean.

We remark on two interesting features that can be observed from inferences of the pressure profiles.
The first is that there are two flat spots near the pedestal top at \(\psi \approx 0.87\) and \(\psi \approx 0.9\), visible in the reduction of the gradients at these locations.
Using \(N_\psi\), which captures the effective number of measurements per time slice used to infer the derivative at a given point, we can see that the flat spots come from inferences which use between 3-4 spatial measurement channels per timeslice which suggests a degree of credibility, especially for the flat spot close to \(\psi \approx 0.9\).

In fact, the locations of these flat spots approximately line up with the \(q=4/1\) and \(q=9/2\) surfaces, suggesting the presence of additional transport mechanisms associated with these rational surfaces.
The approximate uncertainty of the rational surface locations is shown with the orange band in figure~\ref{fig:elm_fit}.
Note the \(q=4/1\) rational surface location is somewhat further off from the location of the flat spot.
This could be a result of the lower credibility of the flat spot location due to the reduced \(N_{eff}\) used to infer the gradient at that point, or could be a result of systematic uncertainties in the equilibrium reconstruction.
We remark that these reductions in the gradient are also visible in the left-most electron temperature profile in figure~\ref{fig:pedestal}.

The second interesting feature is the inward motion of the peak of the pedestal pressure gradient as the pedestal profile evolves during the inter-ELM cycle.
In \cite{nelson_time-dependent_2021}, this behavior was linked to up-chirping of magnetic fluctuations observed in the pedestal region through the predicted dependence of the MTM frequency on the peak value of \(\omega_*\).
We repeat the time-dependent MTM analysis using the profiles inferred from GPR.
We plot the evolution of the peak of \(T_e \frac{\partial(\log p_e)}{\partial\psi} \propto\omega_*\), as well as the location of the peak of \(\omega_*\) in figure~\ref{fig:elm_evolution}.
Consistent with the magnetic fluctuations presented in figure 5 of \cite{nelson_time-dependent_2021}, we observe that the peak of \(\omega_*\) has an up-chirping behavior in the first 30-40 ms of the inter-ELM cycle, followed by a flattening.
Moreover, we observe that the peak of \(\omega_*\) starts unaligned with the \(q=26/5\) surface, before aligning with it later in the inter-ELM cycle.

\section{Summary and Outlook} \label{sec:summary}

In summary, this work has developed two tools, a low-pass filter analogy and \(N_{eff}\), to quantify resolution limits for the inference of differentiable profiles using GPR.
These tools can be used to replace trial-and-error in the fitting or hyperprior development process with quantitative metrics that can be used to engineer the tradeoff between over-fitting and over-regularization.
We summarize the main practical takeaways from the analyses:
\begin{itemize}
    \item For finitely sampled data, GPR does not infer the `true' ensemble averaged profile from the data, but rather a low-pass filtered version of this profile.
    Examples in section \ref{sec:d3d} show how over-regularization can reduce gradients on the order of \(\sim10\%\) compared to properly regularized profiles.
    Given the tendency for profiles to lie near marginal stability boundaries, this low-pass filtering could impact the predicted transport or the stability of different modes.
    \item In the case of regularly sampled data, the transfer function associated with the low-pass filtering applied by GPR has an analytic expression \eqref{eq:ctsfilter} that can be used to `design' the GPR inference.
    For the Mat\'{e}rn family of kernels, which includes the popular squared exponential kernel as a limiting case, the hyperparameters can be combined into the key dimensionless quantity \(S=\sigma^2 \ell / \sigma_\varepsilon^2\Delta x\), interpreted as the \textit{signal-to-noise rate}.
    Contrary to the intuitive idea that \(\ell\) constrains the length scale of the inference, the effective low-pass filter cutoff frequency \(\xi_*\) depends on both the length scale \(\ell\) as well as \(S\) as seen from \eqref{eq:cutoff}.
    \item A procedure for determining estimates or bounds on hyperparameters is given in section \ref{subsec:lowpass_recomenndations}.
    For inferring profiles with rapid changes in the derivative, the squared exponential kernel may be vulnerable to `ringing' artifacts associated with the Gibbs phenomenon; the Mat\'{e}rn kernel with \(\nu=2.5\) can be considered as an alternative.
    To avoid over-regularization, \(\ell\) and \(S\) should be chosen so that \(\xi_* w_{ped} \gtrsim 1\), where \(w_{ped}\) is the full width at half maximum of the inferred pedestal profile.
    \item The \(N_{eff}\) parameter quantifies the number of independent measurements contributing to the inference a given point, which allows for the credibility of GPR to be assessed in regards to over-fitting.
    It does not rely on any simplifying assumptions and can be evaluated for any choice of kernel.
    A higher \(N_{eff}\) indicates a higher robustness to unmodeled errors in the inference reducing the risk of over-fitting, although this usually also implies stronger regularization of the fitted profiles.
    \item By computing low-pass filter cutoffs and \(N_{eff}\) across ranges of hyperparameter space, it is possible to use GPR to compute point estimate inferences with precisely controlled levels of regularization.
    Although the error bars provided by these inferences will only capture epistemic (within-model) uncertainty and not aleatoric (between-model) uncertainty, the inferences do represent unbiased estimates of approximately low-pass filtered versions of the `true' profiles.
    By cross-referencing against physical expectations, e.g. alignment with magnetic safety factor profiles, it is possible to use these metrics to build credibility in the inference of features with increasingly fine spatial and temporal scales.
\end{itemize}

Detailed demonstrations of how this advice can be applied in practice are given in a case study inferring both time-independent and time-dependent pedestal profiles are shown in section \ref{sec:d3d}.
In particular, we demonstrated how analyzing the low-pass filtering associated with GPR gives a quantitative understanding of the smoothing applied by the inference, leading to a better understanding of credible interpretations of pedestal measurements into quantities such as the peak gradient and pedestal width.
We also demonstrated the power of GPR to credibly infer time-dependent evolution of pedestal properties over an inter-ELM cycle while simultaneously inferring subtle features such as `flat spots' associated with low-order rational surfaces.

Finally, we remark on current limitations and possible extensions of this work.
The analysis here was primarily done assuming spatially localized measurements with a known, independent Gaussian noise.
Both of these assumptions can be relaxed somewhat.
As mentioned in the introduction, several methods for GPR tomography and integrating data from multiple heterogeneous diagnostic sources exist.
Furthermore, correlated measurement error can be directly encoded into the covariance matrix \(\boldsymbol{\Sigma}_\varepsilon\), and there exist methods to infer noise levels from measurements \cite{mathews_quantifying_2021}.
We also remark that we have primarily explored GPR without the addition of constraints, such as monotonicity or non-negativity of the inferred profiles.
Different methods have been proposed for imposing these constraints, such as monotonicity promoting priors \cite{michoski_gaussian_2024} or enforcing linear inequality constraints \cite{da_veiga_gaussian_2012}.
As long as an explicit linear or affine relationship between the measurements and the inferred profile can be computed, the equivalent kernel analysis can likely be extended to these cases to derive similar quantities like the cutoff frequency and \(N_{eff}\).

\ack

This material is based upon work supported by the U.S. Department of Energy, Office of Science, Office of Fusion Energy Sciences, using the DIII-D National Fusion Facility, a DOE Office of Science user facility, under Award(s) DE-SC0022164, DE-FG02-04ER54742, DE-FC02-04ER54698, and DE-SC0022270.

\paragraph{Disclaimer}
This report was prepared as an account of work sponsored by an agency of the United States Government. Neither the United States Government nor any agency thereof, nor any of their employees, makes any warranty, express or implied, or assumes any legal liability or responsibility for the accuracy, completeness, or usefulness of any information, apparatus, product, or process disclosed, or represents that its use would not infringe privately owned rights. Reference herein to any specific commercial product, process, or service by trade name, trademark, manufacturer, or otherwise does not necessarily constitute or imply its endorsement, recommendation, or favoring by the United States Government or any agency thereof. The views and opinions of authors expressed herein do not necessarily state or reflect those of the United States Government or any agency thereof.

\section*{Data Availability Statement}
The code associated with this work can be found at https://github.com/Maplenormandy/gpr-resolution

\appendix

\section{Derivation of Low-Pass Filter}\label{app:lowpass_derivation}

For a bi-infinite sequence \(g_j\), define the discrete Fourier transform
\begin{equation*}
    \mathring{\mathcal{F}}[g_j](\xi \Delta x) = \sum_{j=-\infty}^{\infty} g_j e^{-\mathfrak{i}2\pi \xi \Delta x j}
\end{equation*}
Note the continuous and discrete Fourier transforms are related \cite{proakis_digital_1996} by
\begin{equation*}
    \mathring{\mathcal{F}}[f(x_i)](\xi \Delta x) = \frac{1}{\Delta x} \sum_{j=-\infty}^{\infty}\mathcal{F}[f](\xi + \tfrac{j}{\Delta x})
\end{equation*}

For the bi-infinite sequence of measurements \(Y_i = f(i \Delta x) + \varepsilon_i\), equation \eqref{eq:fhat} can be rewritten using an intermediate sequence \(g_j\) via
\begin{gather*}
    \sum_{j=-\infty}^{\infty} [\kappa(x_i - x_j) g_j] + \sigma_\varepsilon^2 g_i = Y_i \\
    \hat{f}(x) = \sum_{j=-\infty}^{\infty} \kappa(x-x_j) g_j
\end{gather*}
Notice that \(\bar{f}(x)\) can be written in terms of \(\bar{g}_j := \mathbb{E}[g_j | f]\),
\begin{subequations}
    \begin{gather}
        \sum_{j=-\infty}^{\infty} [\kappa(x_i - x_j) \bar{g}_j] + \sigma_\varepsilon^2 \bar{g}_i = f(x_i) \label{eq:conv1} \\
        \bar{f}(x) = \sum_{j=-\infty}^{\infty} \kappa(x - x_j) \bar{g}_j
        \label{eq:conv2}
    \end{gather}
\end{subequations}
Taking the Fourier transform of \eqref{eq:conv2} and using the discrete convolution formula on \eqref{eq:conv1} results in the explicit formula
\begin{equation} \label{eq:filter}
    \mathcal{F}[\bar{f}](\xi) = \frac{\mathcal{F}[\kappa](\xi)}{\sigma_\varepsilon^2 + \mathring{\mathcal{F}}[\kappa(x_i)](\xi \Delta x)} \mathring{\mathcal{F}}[f(x_i)](\xi \Delta x)
\end{equation}

Now, when \(\xi \Delta x \lesssim 1\) and \(\mathcal{F}[f],\mathcal{F}[\kappa]\) decay quickly enough (the latter can be guaranteed if \(\Delta x/\ell \ll 1\)),
\begin{equation*}
    \mathring{\mathcal{F}}[f(x_i)](\xi \Delta x) \approx \frac{1}{\Delta x} \mathcal{F}[f](\xi)
\end{equation*}
and similar for \(\mathcal{F}[\kappa]\).
Using this approximation on \eqref{eq:filter} results in \eqref{eq:ctsfilter}.

To extend this result to the multivariate case, consider the setup described in section \ref{subsec:filter_multivariate}.
Let \(\mathring{\kappa}\) denote a function \(\kappa\) evaluated on the lattice points.
Then, a direct application of the multivariate convolution theorem shows that
\begin{equation} \label{eq:nfilter}
    \mathcal{F}[\bar{f}](\xi) = \frac{\mathcal{F}[\kappa](\xi)}{\sigma_\varepsilon^2 + \mathring{\mathcal{F}}[\mathring{\kappa}](\xi \Delta x)} \mathring{\mathcal{F}}[\mathring{f}](\xi \Delta x)
\end{equation}

In the case of a direct product kernel
\begin{equation*}
    \kappa(x-x') = \sigma^2 \prod_{r=1}^{n}\kappa_{(r)}(x_{(r)} - x_{(r)}')
\end{equation*}
the continuous and discrete Fourier transform of \(\kappa\) can be explicitly written out as
\begin{equation*}
\begin{gathered}
    \mathcal{F}[\kappa](\xi) = \sigma^2 \prod_{r=1}^{n} \mathcal{F}_{(r)}[\kappa_{(r)}](\xi_{(r)}) \\
    \mathring{\mathcal{F}}[\mathring{\kappa}](\xi \Delta x) = \sigma^2 \prod_{r=1}^{n} \mathcal{F}_{(r)}[\kappa_{(r)}](\xi_{(r)} \Delta x_{(r)})
\end{gathered}
\end{equation*}

\section{Derivation of \(N_{eff}\)} \label{app:neff_derivation}

Fix some given \(x\).
We consider the case of \(\mu = 0\); the case of \(\mu \neq 0\) can be reduced to this case by making the replacements \(\mathbf{Y} \mapsto \mathbf{Y} + \mu(\mathbf{x})\) and \(\hat{f}(x) \mapsto \hat{f}(x) + \mu(x)\).
We can view GPR as an estimation problem for the linear combination
\begin{equation*}
    \bar{f}(x) = \boldsymbol{\beta}(x)^T f(\mathbf{x})
\end{equation*}
using the estimator \eqref{eq:fhat}
\begin{equation*}
    \hat{f}(x) = \boldsymbol{\beta}(x)^T \mathbf{Y}
\end{equation*}
Using the fact that \(\mathbb{E}[\mathbf{Y}|f] = f(\mathbf{x})\), we have that \(\mathbb{E}[\hat{f}(x)|f] = \bar{f}(x)\).

Assume that \(\boldsymbol{\varepsilon} \sim \mathcal{N}(\mathbf{0}, \boldsymbol{\Sigma}_{\varepsilon})\) is distributed as a multivariate Gaussian random variable with zero mean and covariance matrix \(\boldsymbol{\Sigma}_{\varepsilon}\).
We assume \(\boldsymbol{\Sigma}_{\varepsilon}\) is non-singular, although it could be non-diagonal, encoding correlations between the measurement errors.

First, define an inner product \(\ev{\mathbf{a},\mathbf{b}}_{\boldsymbol{\Sigma}_{\varepsilon}} := \mathbf{a}^T \boldsymbol{\Sigma}_{\varepsilon} \mathbf{b}\), and let \(\sigma_{\beta}^{2} := \ev{\boldsymbol{\beta}(x),\boldsymbol{\beta}(x)}_{\boldsymbol{\Sigma}_{\varepsilon}}\).
Since \(\hat{f}(x)|f \sim \mathcal{N}(\bar{f}(x), \sigma_\beta^2)\),
\begin{equation}
    \mathcal{I}_{\hat{f}(x)}[\bar{f}(x)] = \frac{1}{\sigma_\beta^2}
\end{equation}
Note that in the case that \(\boldsymbol{\Sigma}_{\varepsilon}\) is diagonal with elements \(\sigma_i^2\) along the diagonal, then \(\sigma_\beta^2 = \sum_{k=1}^{N} \sigma_k^2 \beta_k(x)^2\).

Next, since \(\mathbf{Y} = f(\mathbf{x}) + \boldsymbol{\varepsilon}\), the conditional random variable \(\mathbf{Y} | f \sim \mathcal{N}(f(\mathbf{x}),\boldsymbol{\Sigma}_{\varepsilon})\), so using the formula for the Fisher information of a Gaussian random variable with respect to its mean,
\begin{equation*}
    \mathcal{I}_{\mathbf{Y}}[f(\mathbf{x})] = \boldsymbol{\Sigma}_{\varepsilon}^{-1}
\end{equation*}
Note here we treat \(f(\mathbf{x})\) as an \(N\)-dimensional vector of parameters.

To reveal the additive structure of this Fisher information, let \(\mathbf{v}_k\) be the orthonormal (in the standard inner product) basis vectors of \(\boldsymbol{\Sigma}_{\varepsilon}\) with corresponding eigenvalues \(\lambda_k^2\).
In this case the random variables \(Z_k := \mathbf{v}_k^T \mathbf{Y}\) are independent, and one can show
\begin{equation*}
\begin{gathered}
    \mathcal{I}_{\mathbf{Y}}[f(\mathbf{x})] = \sum_{k=1}^{N} \mathcal{I}_{Z_k}[f(\mathbf{x})] \\
    \mathcal{I}_{Z_k}[f(\mathbf{x})] = \frac{\mathbf{v}_k \mathbf{v}_k^T}{\lambda_k^2}
\end{gathered}
\end{equation*}
Note that in the case where \(\boldsymbol{\Sigma}_{\varepsilon}\) is diagonal so the measurement errors are uncorrelated, one can take \(Z_k = Y_k\) and \(\lambda_k^2 = \sigma_k^2\).

To compute the Fisher information of \(\mathbf{Y}\) with respect to \(\bar{f}(x)\), which is a linear combination of the parameters \(f(\mathbf{x})\), we need to marginalize out the other \(N-1\) degrees of freedom from the Fisher information matrix.
This can be accomplished by finding a transformation of our parameters \(f(\mathbf{x})\) such that the information matrix with respect to the new parameter coordinates will be diagonal \cite{barndorff-nielsen_inference_2017}.

This can be accomplished by extending \(\boldsymbol{\beta}(x)\) into a basis \(\boldsymbol{\beta}_1 = \boldsymbol{\beta}(x), \boldsymbol{\beta}_2, ..., \boldsymbol{\beta}_N\) which is orthogonal under the inner product 
This can be done for any non-zero \(\boldsymbol{\beta}(x)\) via Gram-Schmidt.
Furthermore, for convenience we can normalize the nuisance parameters such that \(\ev{\boldsymbol{\beta}_j,\boldsymbol{\beta}_j}_{\boldsymbol{\Sigma}_{\varepsilon}} = \ev{\boldsymbol{\beta}_1,\boldsymbol{\beta}_1}_{\boldsymbol{\Sigma}_{\varepsilon}} = \sigma_\beta^2\).

Define a matrix
\begin{equation*}
    B := \begin{bmatrix}
        \boldsymbol{\beta}_1 & \boldsymbol{\beta}_2 & \cdots & \boldsymbol{\beta}_N
    \end{bmatrix}
\end{equation*}
Due to the orthogonality property, \(B^T \boldsymbol{\Sigma}_{\varepsilon} B = \sigma_\beta^2 I\), so
\begin{equation*}
    B^{-1} = \frac{B^T \boldsymbol{\Sigma}_{\varepsilon}}{\sigma_\beta^2}
\end{equation*}

Write \(\bar{\mathbf{f}} = B^T f(\mathbf{x})\).
Using the reparameterization formula for the Fisher information,
\begin{equation*}
\begin{aligned}
    \mathcal{I}_{\mathbf{Y}}[\bar{f}] = B^{-1} \boldsymbol{\Sigma}_{\varepsilon}^{-1} (B^{-1})^T = \frac{1}{\sigma_\beta^2}I
\end{aligned}
\end{equation*}
which is block diagonal.
In particular,
\begin{equation}
    \mathcal{I}_{\mathbf{Y}}[\bar{f}(x)] = \left[\mathcal{I}_{\mathbf{Y}}[\bar{\mathbf{f}}]\right]_{11} = \frac{1}{\sigma_\beta^2}
\end{equation}
so \(\mathcal{I}_{\hat{f}(x)}[\bar{f}(x)] = \mathcal{I}_{\mathbf{Y}}[\bar{f}(x)]\).

Now, to get the information from each independent variable \(Z_k\),
\begin{align*}
    \mathcal{I}_{Z_k}[\bar{\mathbf{f}}] &= B^{-1} \frac{\mathbf{v}_k \mathbf{v}_k^T}{\lambda_k^2} (B^{-1})^T \\
    &= \frac{1}{\sigma_\beta^4 \lambda_k^2} B^T \boldsymbol{\Sigma}_{\varepsilon} \mathbf{v}_k \mathbf{v}_k^T \boldsymbol{\Sigma}_{\varepsilon} B \\
    &= \frac{\lambda_k^2}{\sigma_\beta^4} (\mathbf{v}_k^T B)^T \mathbf{v}_k^T B
\end{align*}
then, if we take the first component to marginalize (note that the matrix is not block diagonal)
\begin{equation*}
    \mathcal{I}_{Z_k}[\bar{f}(x)] = \left[\mathcal{I}_{Z_k}[\bar{\mathbf{f}}]\right]_{11} = \frac{\lambda_k^2 (\mathbf{v}_k^T \boldsymbol{\beta}(x))^2}{\sigma_\beta^4}
\end{equation*}
Note that if \(\boldsymbol{\Sigma}_{\varepsilon}\) is diagonal and we take \(Z_k = Y_k\) as before, then
\begin{equation*}
    \mathcal{I}_{Y_i}[\bar{f}(x)] = \frac{\sigma_i^2 \beta_i(x)^2}{\left(\sum_{k=1}^{N} \sigma_k^2 \beta_k(x)^2\right)^2}
\end{equation*}
as desired.

\section{Derivation of \(N_{eff}\) Bound}\label{app:neff_inequality}

For convenience, we reiterate the formula
\begin{equation} \label{eq:f_dev}
    \tilde{f}(x) - \hat{f}(x) = \sum_{j=1}^{M} \left[ \delta_j \sum_{i\in P_j} \beta_i(x) \sigma_i \right]
\end{equation}
and notice that the derivative satisfies a similar equation
\begin{equation*}
    \tilde{f}'(x) - \hat{f}'(x) = \sum_{j=1}^{M} \left[ \delta_j \sum_{i\in P_j} \beta_i'(x) \sigma_i \right]
\end{equation*}

Inequality \eqref{eq:neff_inequality} is derived by the following steps,
\begin{align*}
    \operatorname{Var}[\tilde{f}(x)-\hat{f}(x) | f]^2
    &= \left( \sum_{j=1}^{M} \operatorname{Var}[\delta_j] \left(\sum_{i\in P_j} \sigma_i \beta_i(x)\right)^2 \right)^2 \\
    &\le \left( \sum_{j=1}^{M} |P_j| \operatorname{Var}[\delta_j]  \sum_{i\in P_j} \sigma_i^2 \beta_i(x)^2 \right)^2 \\
    &\le \left(\sum_{j=1}^{M} |P_j|^2 \operatorname{Var}[\delta_j]^2\right) \left(\sum_{j=1}^{M} \left(\sum_{i\in P_j} \sigma_i^2 \beta_i(x)^2\right)^2\right) \\
    &= \overline{D^2} \frac{\sigma_\varepsilon^2(x)^2}{N_{\mathbf{P}}[\bar{f}(x)]}
\end{align*}
where dividing both sides by \(\sigma_{\varepsilon}^2(x)^2\) gives the desired result.

To explain this sequence of steps, the first line uses the fact that the \(\delta_j\) are all independent, and hence the variance of the sum \eqref{eq:f_dev} is the sum of the variances, and also uses \(\operatorname{Var}[\alpha X] = \alpha^2 \operatorname{Var}[X]\) for constant \(\alpha\).
The second line uses the relationship between the arithmetic mean and the root mean square,
\begin{equation*}
    \left(\sum_{i\in P_j} \sigma_i \beta_i(x)\right)^2 \le |P_j| \sum_{i\in P_j} \sigma_i^2 \beta_i(x)^2
\end{equation*}
where equality holds if all of the \(\sigma_i \beta_i(x)\) are equal for \(i \in P_j\).

The third line comes from an application of the Cauchy-Schwarz inequality \((\mathbf{a} \cdot \mathbf{b})^2 \le \norm{a}^2 \norm{b}^2\) on two vectors \(\mathbf{a},\mathbf{b}\) which have components \(a_j = |P_j| \operatorname{Var}[\delta_j]\) and \(b_j = \sum_{i\in P_j} \sigma_i^2 \beta_i(x)^2\).
Notice that \(\mathbf{a}\cdot\mathbf{b} = \norm{a} \norm{b} \cos(\theta)\) where \(\theta\) is the angle between \(\mathbf{a}\) and \(\mathbf{b}\), so equality holds when \(\mathbf{a} \parallel \mathbf{b}\).

The fourth and final line comes from
\begin{equation*}
    \frac{1}{N_{\mathbf{P}}[\bar{f}(x)]} = \frac{\sum_{j=1}^{M} \left(\sum_{i\in P_j} \sigma_i^2 \beta_i(x)^2\right)^2}{\left(\sum_{k=1}^{N} \sigma_k^2 \beta_k(x)^2\right)^2}
\end{equation*}
Finally, notice that if \(\beta_i(x)\) is replaced by \(\beta_i'(x)\) in each line of the derivation, an exactly analogous result is derived for \(\tilde{f}'(x) - \hat{f}'(x)\).

From the discussion earlier, we can show that the worst-case scenario predicted by the bound can happen in the following scenario:
Suppose for each bin \(P_j\), each measurement \(i \in P_j\) has the same error bar \(\sigma_i\) and measurement location \(x_i\), which leads to all of the \(\sigma_i \beta_i(x)\) being equal in the bin and hence equality on the second line.
Then, the components \(b_j = |P_j| \sigma_i^2 \beta_i(x)^2\), so if \(\operatorname{Var}[\delta_j] \propto \sigma_i^2 \beta_i(x)^2\) for one of the \(i \in P_j\), then equality holds on the third line.
This achieves the worst-case deviation predicted by the bound.
Intuitively, this represents the case where the variance of the unmodeled error \(\delta_j\) is proportional to the squared sensitivity of the inference to the measurements in the bin \(P_j\).

\section*{References}
\bibliography{main}
\bibliographystyle{iopart-num}

\end{document}